\documentclass[apl,tightenlines,superscriptaddress]{revtex4-2}

\usepackage[toc,page]{appendix}
\usepackage{amssymb,mathtools,physics,amsfonts}

\usepackage{subcaption}
\usepackage{float}
\usepackage{multirow}

\usepackage{amsthm}
\usepackage{amsmath}
\usepackage{bbm, dsfont}
\usepackage{graphicx}
\usepackage[usenames,dvipsnames]{color}
\usepackage[colorlinks=true,citecolor=magenta,urlcolor=blue]{hyperref}
\usepackage[table]{xcolor}
\usepackage{colortbl}
\usepackage{color}
\usepackage{multirow}
\usepackage{hhline}
\usepackage{booktabs}
\usepackage[normalem]{ulem}
\usepackage{siunitx}
\usepackage{enumitem}
\usepackage{braket}
\usepackage{siunitx}
\usepackage{textcomp}
\usepackage{booktabs}
\usepackage[textwidth=3cm, textsize=footnotesize]{todonotes}

\usepackage[capitalise]{cleveref}

\DeclareUnicodeCharacter{2009}{\,}

\DeclareSIUnit\permille{\text{\textperthousand}}

\graphicspath{{./Figures/}}

\newcolumntype{C}[1]{ >{ \centering\arraybackslash}p{#1}}

\definecolor{gray4}{gray}{0.8}
\definecolor{gray2}{gray}{0.6}

\newcommand{\quotes}[1]{``#1''}

\newcommand{\affilationVigoOne}{Vigo Quantum Communication Center, University of Vigo, Vigo E-36310, Spain}
\newcommand{\affilationVigoTwo}{Escuela de Ingeniería de Telecomunicación, Department of Signal Theory and Communications, University of Vigo, Vigo E-36310, Spain}
\newcommand{\affilationVigoThree}{AtlanTTic Research Center, University of Vigo, Vigo E-36310, Spain}
\newcommand{\affilationGeneva}{Department of Applied Physics, University of Geneva, 1205 Geneva, Switzerland}

\newcommand{\decoyrefs}{\cite{hwang_quantum_2003,lo_decoy_2005,wang_beating_2005}}

\begin{document}

\title{Simplified quantum key distribution implementation secure in the presence of state preparation flaws}	

\author{Ainhoa Agulleiro\footnote{These authors contributed equally to this work.}}\email{aagulleiro@vqcc.uvigo.es}
	\affiliation{\affilationVigoOne}
	\affiliation{\affilationVigoTwo}
	\affiliation{\affilationVigoThree}

\author{Fadri Gr\"unenfelder\footnotemark[1] \footnote{Now at \affilationGeneva.}}
	\affiliation{\affilationVigoOne}
	\affiliation{\affilationVigoTwo}
	\affiliation{\affilationVigoThree}

\author{Rapha\"el Houlmann}
	\affiliation{\affilationGeneva}

\author{Ana Blázquez}
	\affiliation{\affilationVigoOne}
	\affiliation{\affilationVigoTwo}
	\affiliation{\affilationVigoThree}

\author{Hugo Zbinden}
	\affiliation{\affilationVigoOne}
	\affiliation{\affilationVigoTwo}
	\affiliation{\affilationVigoThree}
	
\author{Davide Rusca}
	\affiliation{\affilationVigoOne}
	\affiliation{\affilationVigoTwo}
	\affiliation{\affilationVigoThree}
	
\date{\today}


\begin{abstract}
    We present an implementation of a three-state BB84 protocol with time-bin encoding, one decoy state and a simplified measurement scheme that uses passive basis choice. Our system simplifies the state characterization with respect to previous iterations. We also adapt the loss-tolerant method to our protocol, thus dealing with the measured state preparation flaws. We compare the obtained phase error rate and secret key rate when including the state imperfections and when assuming perfect states. Our results highlight the importance of characterization and implementation security.
\end{abstract}

\maketitle

\section{Introduction}

Quantum key distribution (QKD) promises information-theoretically secure communications when combined with the one-time pad. Since the proposal of the BB84 protocol \cite{brassard1984quantum}, many technical advances have been made that have transformed QKD into a well-established technology. For example, the decoy-state method \decoyrefs  allows the use of faint laser pulses instead of single-photons, which makes QKD much more practical. Nowadays, state-of-the-art experiments have reached very high key rates \cite{grunenfelder_fast_2023,li_high-rate_2023}. However, pushing the performance comes at a cost, since the higher the speed, the harder it becomes for Alice to reliably create her states \cite{diamanti_practical_2016}. This leads to state preparation flaws (SPFs) and, more concerningly, a variety of different side-channels like intensity, phase and bit-and-basis correlations \cite{yoshino_quantum_2018,grunenfelder_performance_2020}. Implementation security aims to include all these imperfections in the theory, such that even an imperfect system can be provably secure \cite{sun_review_2022,zapatero_implementation_2025}. For example, the loss-tolerant (LT) method proposed by Tamaki \textit{et al.} includes SPFs \cite{tamaki_loss-tolerant_2014} and has been successfully implemented in works like \cite{xu_experimental_2015}, while more recent security proofs aim for security in the presence of correlations \cite{pereira_quantum_2019,pereira_quantum_2020,zapatero_security_2021,sixto_security_2022,curras-lorenzo_security_2023,curras-lorenzo_security_2025}.

In this context, we present an implementation of a three-state BB84 protocol with a simplified measurement scheme \cite{rusca_security_2018}. Previous implementations of this protocol have been shown to achieve secret key rates of around 60Mbps \cite{grunenfelder_fast_2023} and to be able to distribute keys over a channel of more than 421km of ultra-low loss fiber\cite{boaron_secure_2018}.
While these works proved the practicality and capability for high performance of this protocol, they did so while making assumptions on the state preparation that were not verified.
The security proof by Rusca \textit{et al.} considers perfect state preparation, which means that if the states have any kind of imperfection as it is common in high-speed implementations, the security can be compromised.
In this work, we introduce several changes to ensure security even in the presence of SPFs.
On the experimental side, we limit the repetition rate to avoid overlap between time bins.
This makes the system slower but it allows us to measure the SPFs without changing the setup.
On the theoretical side, we change the security analysis to account for SPFs.
To do so, instead of assuming that the state preparation is perfect as done in \cite{rusca_security_2018}, we adapt the LT method to our protocol. In this way, we are able to successfully distribute secret keys secure against collective attacks over 150km  of fiber without making the perfect state preparation assumption. 

The structure of this paper is as follows.
In \cref{section:protocol,section:setup}, we describe the QKD protocol and how it is implemented in the experiment. In \cref{section:state-characterization}, we describe the state characterization method and the results that it yields. In \cref{section:security proof}, we briefly outline the new security proof. In \cref{section:secret-key-exchange}, we show the results of the secret key exchange, for which we use the results presented in the previous sections.
Lastly, the conclusions are summarized in \cref{section:discussion}. 
Appendix \ref{appendix:delay_mismatch} contains visibility measurements and explains the effect of interferometer imperfections on the protocol performance.
Appendix \ref{appendix:state characterization} shows calculations regarding the state characterization and contains a discussion about how the characterization can be performed securely. 
Finally, Appendix \ref{appendix:security proof} contains the calculations for the security analysis.


\section{Protocol description}\label{section:protocol}
We implement the simplified three-state BB84 protocol with one decoy presented in
\cite{rusca_security_2018}. 
Alice prepares her states using time bin encoding on phase-randomized weak coherent states.  
She chooses the Z basis with probability $p_Z^A$, for which she can send either an early pulse, $\ket{\sqrt{\mu}}_{t0}\ket{\emptyset}_{t1}$, or a late pulse,  $\ket{\emptyset}_{t0}\ket{\sqrt{\mu}}_{t1}$, with equal probability. 
Here, the subindices $t0$ and $t1$ refer to the time bins, $\mu$ is the mean photon number and $\ket{\emptyset}$ is the vacuum state.
These states are referred to as 0Z and 1Z, since their single-photon components are the eigenstates of the Z basis, denoted as $\ket{0}$ and $\ket{1}$.
On the other hand, only one state is created for the X basis, which we denote with 0X and is given by  $\ket{\sqrt{\mu/2}}_{t0}\ket{\sqrt{\mu/2}}_{t1}$.
Its single-photon component is given by  $\ket{+}=(\ket{0}+\ket{1})/\sqrt{2})$. 
In accordance to the one-decoy state method \cite{rusca_finite-key_2018}, $\mu$ is chosen in each round between the signal ($\mu_0$) and decoy ($\mu_1<\mu_0$) levels with probabilities $p_{\mu0}$ and $1-p_{\mu0}$, respectively.

\begin{figure}[h!]
	\centering
	\includegraphics[width=0.75\linewidth]{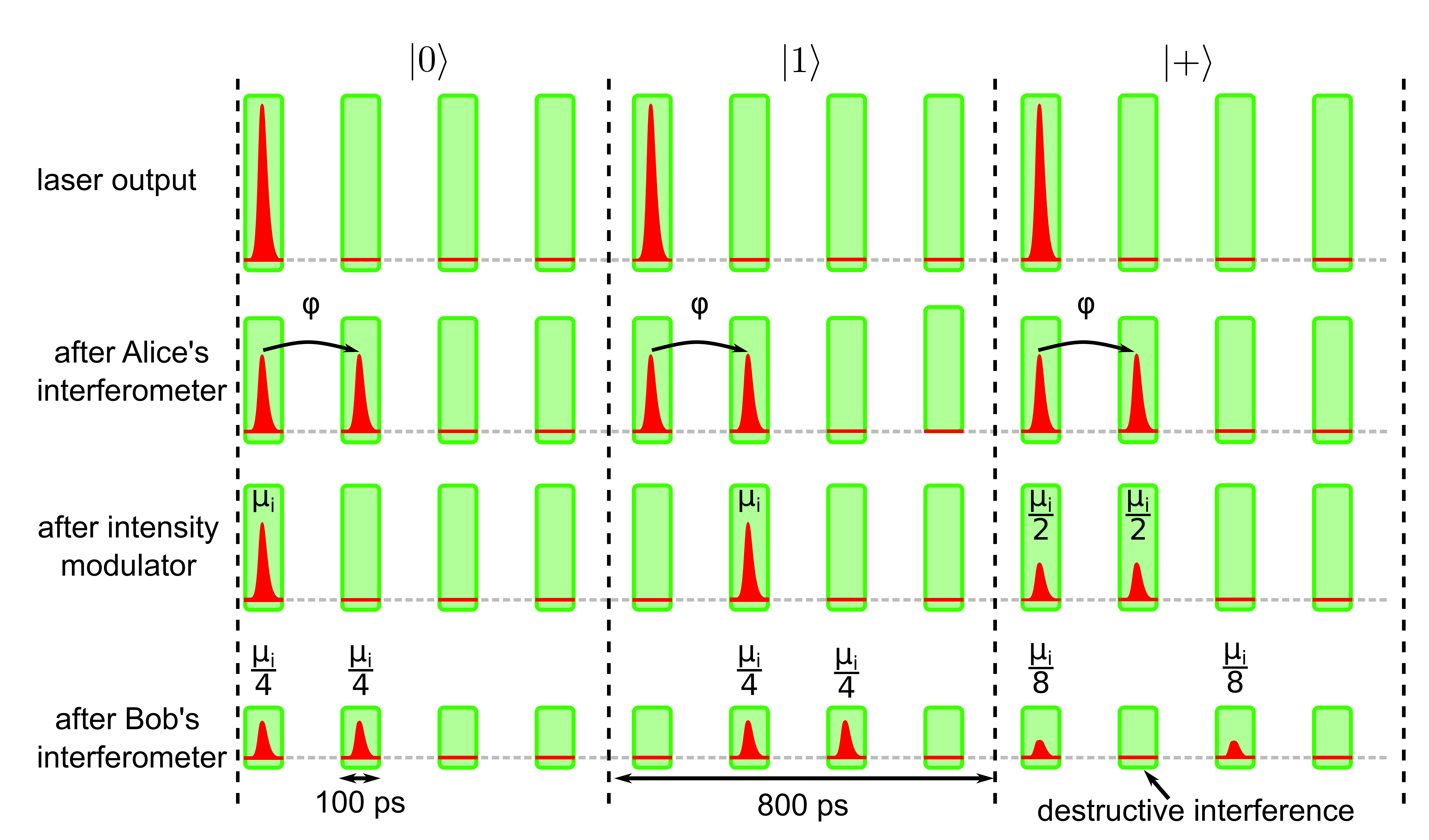}
	\caption{Schematic drawing showing the states at different stages of the protocol when there is no interference from Eve nor imperfections in the transmitter and receiver.}
	\label{fig:states}
\end{figure}

Bob measures the incoming states in the Z (X) basis with probability $p_Z^{B}$ ($p_X^B=1-p_Z^B$). In the Z basis, he measures the time of arrival. In the X basis, he makes the early and late time bins interfere. His detections are denoted as Z0, Z1 for the early and late time bins of the Z basis and X0, X1 and X2 for the early, middle and late time bins in the X basis. The detections in X1 are used to monitor the relative phase. The side bins X0 and X2, on the other hand, are equivalent to a Z basis measurement (see \cref{fig:states}). The rounds in which Alice sends a Z state and Bob gets a detection in the same basis are used for key generation. The rest of the events are used for parameter estimation. 

\begin{figure}[h!]
	\centering
	\includegraphics[width=\linewidth]{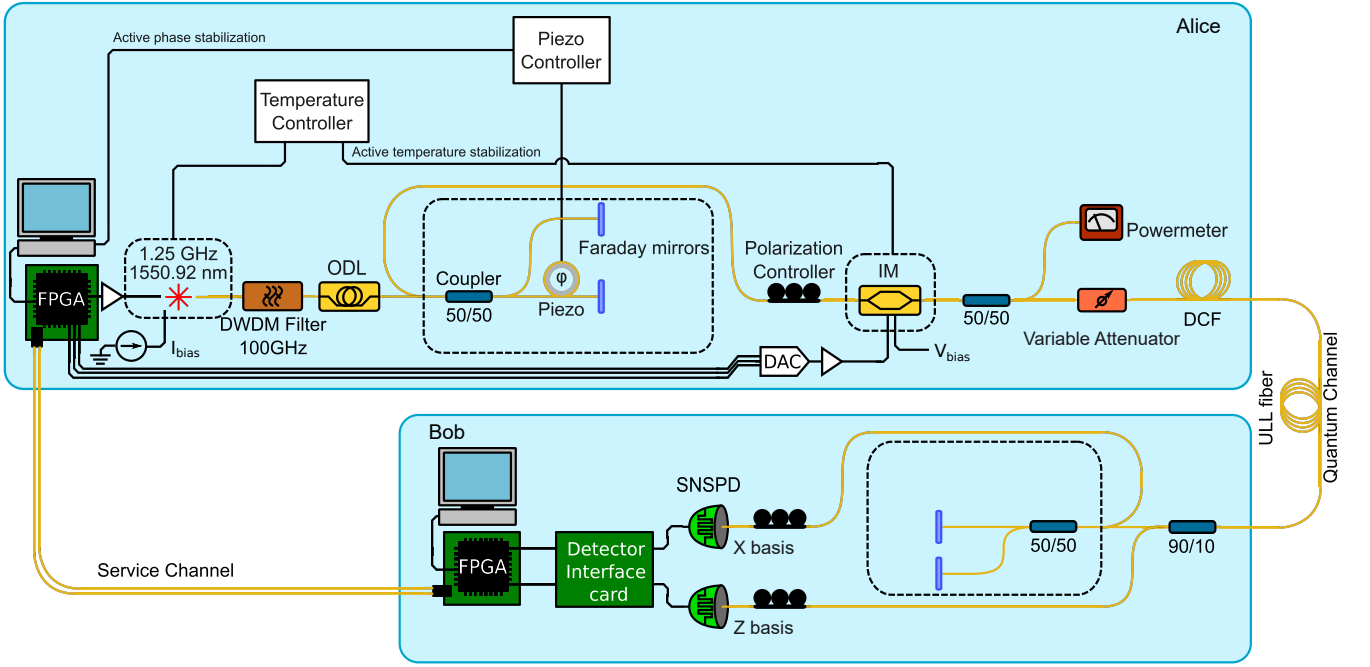}
	\caption{Schematic representation of the experimental setup. The black lines indicate electrical connections. The yellow lines represent single-mode optical fiber. The boxes with a dashed outline indicate thermally isolated environments. The boxes containing the laser and IM have active temperature stabilization. However, the boxes containing the interferometers are not temperature controlled. Instead, their relative phase is stabilized through the action of the piezoelectric element in Alice's interferometer. FPGA: field-programmable gate array. DWDM: dense wavelength division multiplexer. ODL: optical delay line. IM: intensity modulator. DAC: Digital to analogue converter. DCF: dispersion compensation fiber. ULL: ultra-low loss. SNSPD: superconducting nanowire single-photon detector.}
	\label{fig:setup}
\end{figure}

\section{Implementation}\label{section:setup}
The experimental setup used for the implementation of this protocol is shown in \cref{fig:setup}.
On Alice's side, a distributed feedback laser with a modulation bandwidth of 10 GHz (Gooch and Housego AA0701) is gain-switched with a 1.25GHz clock coming from her FPGA. 
The signal is amplified and a bias current is added. 
The resulting optical pulses go through a dense wavelength division multiplexer (DWDM) acting as a spectral filter centered at 1550.92nm. 
This reduces the spectrum of the chirped pulses, which as a side effect decreases their width. 
The amplifier parameters, current bias and filter are chosen to minimize the full width at half maximum (FWHM) of the laser pulses, which is 37.5 ps. 
Each laser pulse is split into two coherent pulses separated by 202 ps through the action of a Michelson interferometer with an equivalent path difference. The interferometer consists of a 50:50 beam-splitter and two Faraday mirrors, which makes it polarization independent. 
After the interferometer, an intensity modulator (IM) of 10GHz bandwidth (Exail MXR-LN-10) modulates the pairs of pulses to create the desired states. 
On the one hand, it chooses whether the early, late or both pulses (with half the intensity) go through, thus determining the basis and bit of the qubit (see \cref{fig:states}). 
On the other hand, it also chooses the mean photon number of the overall state, thus choosing between the decoy intensity levels ($\mu_0$ and $\mu_1$).
The polarization at the input of the IM is changed by a manual polarization controller (PC) so that the extinction ratio is maximized. 
The electrical signal needed for controlling the IM is created from three bits coming from the FPGA, which are injected into a digital to analogue converter (DAC). 
The analogue signal is amplified and an additional bias voltage is applied by a power supply. 
To choose the states, the FPGA uses a deterministic pseudorandom number generator based on Advanced Encryption Standard (AES), but it could be seeded by a quantum random number generator. 
An optical delay line (ODL) placed just before the interferometer takes care of aligning the laser pulses well inside the intensity modulation windows.
After the IM, a coupler splits the light into a powermeter and a variable optical attenuator (VOA). 
The powermeter is used to monitor the average power of Alice's signal. 
This power is used to calculate which attenuation is needed in the VOA to have the desired mean photon number of the states at the output of Alice. 
Finally, the light goes through dispersion compensation fiber (DCF) to compensate the chromatic dispersion due to the length of the quantum channel (QC). 
The loss of the DCF is also taken into account when calculating the necessary attenuation.

The QC consists of spools of ultra-low loss (ULL) fiber with 0.17dB/km of loss. 
The service channel consists of two 20 m single mode fibers, one for each direction of classical communication.

On Bob's side, a coupler performs passive basis choice, splitting 90\% of the power to the Z basis and 10\% to the X basis.  
In the Z basis, the signals are directly sent to a superconducting nanowire single-photon detector (SNSPD), which measures the time of arrival. 
In the X basis, the pairs of pulses are made to interfere in a Michelson interferometer with 198 ps arm difference. 
The interference is monitored by another SNSPD. 
Both SNSPDs have a detection efficiency of 90\%, timing jitter of around 30ps, dark count rates of 8 counts per second and recovery time of 35ns.
Just before each detector, a PC changes the polarization of the incoming light in order to maximize the efficiency. 
This is done because the SNSPDs have polarization dependent loss. 
An in-house made interface card processes the signals of the detectors to make them compatible with the FPGA of Bob. 
The card has a delay line for each detector, which allows Bob to shift his detection in steps of 5ps.
This is used to align the counts in their time bins during the calibration.
Moreover, the delay lines are also used for an automatic feedback loop that shifts all the time bins simultaneously to compensate the varying length of the ULL fiber due to small temperature changes. 

The relative phase between Alice's and Bob's interferometer is kept constant with a stabilization loop. 
To that end, the qubit error rate in the X basis (QBERX) is monitored and used as feedback.
A voltage is then applied accordingly to a piezoelectric element in Alice's interferometer, which changes the birefringence of the fiber in one arm.
Effectively, this stabilizes the relative phase between the early and late pulses of the 0X state.

The 4ps difference in the delay of Alice's and Bob's interferometers and additional imperfections like intensity imbalance of the arms could impact the performance of the protocol \cite{millet_influence_2025}.
We measured the visibility of the interference in Bob's interferometer of the pairs of coherent pulses generated by Alice's interferometer to quantify these effects (see \cref{appendix:delay_mismatch} for more details).
The visibility measured in this way is of 98\%, while the independent visibilities measured with a laser operated in continuous wave were of 99.8\% and 99.7\% for Alice and Bob, respectively.

Error correction (EC) and privacy amplification (PA) are not implemented. Instead, the quantum bit error rate QBER of the Z basis, QBERZ, and the phase error rate are estimated from the acquired statistics. The leaked bits during EC are calculated from the QBERZ assuming a low density parity check (LDPC) algorithm like the one in \cite{constantin_fpga-based_2017}. This is used alongside the estimated phase error rate to calculate the final SKR. Note that it would still be possible to perform PA and EC in real time with already existing algorithms as proved in state-of-the-art experiments like \cite{grunenfelder_fast_2023,li_high-rate_2023}.

\section{State characterization}\label{section:state-characterization}

Previous works using a simplified three-state protocol with one decoy \cite{boaron_simple_2018-1,boaron_secure_2018,grunenfelder_fast_2023} make use of the efficient time bin configuration proposed by Rusca \textit{et al} \cite{rusca_security_2018}, which allows for a higher repetition rate. In contrast, we leave enough separation between rounds so that there is no such overlap in the Bob's X measurement. The lack of this overlap simplifies the state characterization at the price of a lower state repetition rate. Indeed, we are able to measure the state preparation flaws (SPFs) by simply running the protocol and letting Bob reveal all of his bits to Alice. This is done before the actual secret key exchange, as part of the calibration of the experiment. 
By looking at the obtained statistics of the detections on each time bin conditioned on the preparation of state $A\in\{0Z, 1Z, 0X\}$ with mean photon number $\mu\in\{\mu_0, \mu_1\}$, one can calculate the polar and azimuthal angles in the Bloch sphere of each state, $\theta_{A,\mu}$ and $\varphi_{A, \mu}$, as well as the mean photon numbers $\mu_{A}$ (see \cref{appendix:state characterization} for detailed calculations).
The used characterization method also incorporates the visibility due to the delay mismatch of the interferometers in a natural way, as explained in Appendices \ref{appendix:delay_mismatch} and \ref{appendix:state characterization calculations}.

\begin{figure}[h!]
\centering
\begin{subfigure}{0.45\linewidth}
	\includegraphics[width=\linewidth]{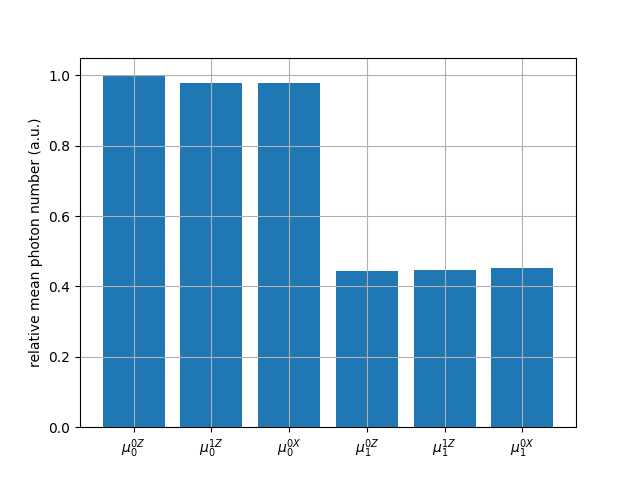}
	\caption{}
	\label{fig:intensities_101km}
\end{subfigure}
\begin{subfigure}{0.45\linewidth}
	\includegraphics[width=\linewidth]{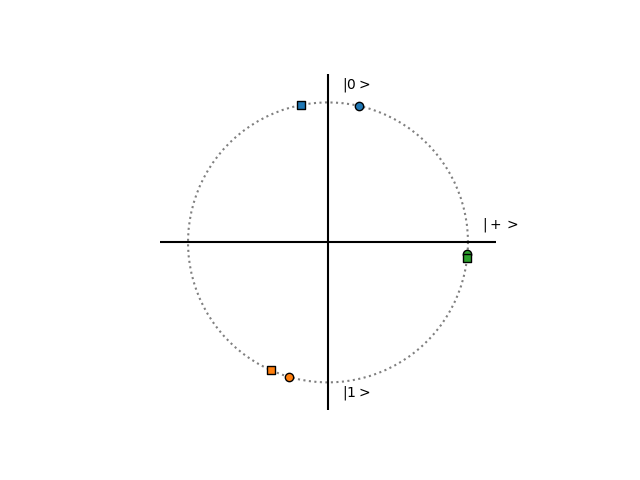}
	\caption{}
	\label{fig:projection_xz_plane}
\end{subfigure}
\caption{(a) Plot of the relative mean photon number of each state using $\mu_0^{0Z}$ as a reference. (b) Projection of the characterized states to the XZ plane of the Bloch sphere. The circles represent the states at the signal level ($\mu_0$), while the squares are the states at the decoy level ($\mu_1$). Both plots are from the statistics using 101.4km QC length.}\label{fig:state characterization}
\end{figure}

One could use different schemes to perform the state characterization, as explained in Appendix \ref{appendix:characterization configurations}. 
In our case, because Alice and Bob are in the same lab, we perform the state characterization directly through the QC.
However, we remark that in a practical setting, the characterization must be done ensuring that Eve cannot affect it through the QC (see \cref{appendix:characterization configurations} for the proposed methods).
We do the characterization for each of the distances used for the key exchange.
We do the measurement for each distance to ensure that the SPFs stayed similar, since not all the key exchanges were carried out on the same day and the fact that the state preparation probabilities were changed meant that the bias voltage of the IM had to also be changed.
Both of these facts could slightly change the states prepared.
We also repeated the measurements after replacing the ULL fiber with a shorter single-mode fiber and a VOA.
The results are shown in \cref{tab:table1}. 
We can see that the SPFs are not significantly increased by the high lengths of fiber. 
This implies that the dispersion was well compensated for.
On the other hand, from the same histograms one can also calculate the ratio between the intensities of the decoy levels for each state (see \cref{fig:intensities_101km}). 
Note that, as shown by \cref{fig:state characterization}, the estimated SPFs of the qubits depend on which decoy level is used and, inversely, the intensities of the decoy levels depend on the qubit choice.  This is so because both the qubit and decoy encoding are performed by the same IM, meaning that memory effects will introduce correlations between them. Similar correlations between qubit encoding and intensity levels have been observed in other QKD transmitters and it is possible to prove security \cite{xing_cross_correlations_2025}. However, in this work we only consider SPFs. For this reason, we only use the state characterization with signal intensity in the secret key exchange using the loss-tolerant method.

\begin{table}[h!]
	\begin{center}
		\caption{State characterization results. The angles of several measurements are averaged using the mean direction and the uncertainties are calculated with the circular standard deviation. ($^{*}$) means that the measurement was done with an optical attenuator between Alice and Bob, not spools of ULL fiber.}
		\label{tab:table1}
		\begin{tabular}{|c|c|c|c|c|c|c|c|}\hline
	      &  & \multicolumn{2}{|c|}{0Z} & \multicolumn{2}{|c|}{1Z} & \multicolumn{2}{|c|}{0X}  \\ \hline
		QC & decoy level & $\theta$ (rad) & $\varphi$ (rad) & $\theta$ (rad) & $\varphi$ (rad) & $\theta$ (rad) & $\varphi$ (rad) \\ \hline
		\multirow{2}{*}{101.4km ULL}& $\mu_0$ & $0.222(9)$ & $\rm 0.5(2)$ & $2.858(9)$ & $\rm 2.7(1)$ & $\rm 1.66(1)$ & $ \rm 0.14(2)$ \\ 
		&$\mu_1$ & $0.192(9)$ & $\rm 1.9(1)$ & $ 2.722(9)$ & $2.6(1)$ & \rm $1.68(1)$ & $\rm 0.16(2)$ \\ \hline
		\multirow{2}{*}{112.6km ULL}& $\mu_0$ & $0.28(1)$ & $\rm 0.6(1)$ & $2.85(1)$ & $\rm 2.4(1)$ & $\rm 1.63(2)$ & $ \rm 0.16(2)$ \\ 
		&$\mu_1$ & $0.25(1)$ & $\rm 1.61(9)$ & $2.72(1)$ & $2.38(9)$ & \rm $1.66(2)$ & $\rm 0.20(3)$ \\ \hline
		\multirow{2}{*}{151.0km ULL}& $\mu_0$ & $0.23(2)$ & $\rm 0.5(2)$ & $2.88(2)$ & $\rm 2.6(2)$ & $\rm 1.66(3)$ & $ \rm 0.12(4)$ \\ 
		&$\mu_1$ & $0.15(3)$ & $\rm 2.1(3)$ & $2.75(3)$ & $2.6(2)$ & \rm $1.69(3)$ & $\rm 0.13(5)$ \\ \hline
        \multirow{2}{*}{25 dB attenuation $^{*}$}& $\mu_0$ & $0.18(2)$ & $\rm 0.7(2)$ & $2.86(2)$ & $\rm 2.6(2)$ & $\rm 1.67(3)$ & $ \rm 0.12(4)$ \\ 
		&$\mu_1$ & $ 0.21(2)$ & $\rm 1.9(2)$ & $2.73(3)$ & $2.6(2)$ & \rm $1.69(3)$ & $\rm 0.15(5)$ \\ \hline
		\end{tabular}
	\end{center}
\end{table}

\newpage
\section{Security analysis}\label{section:security proof}

In order to use the measured SPFs in the security analysis, we adapt the LT method \cite{tamaki_loss-tolerant_2014} to our scheme.
To contextualize our results, let us first briefly introduce the LT method.
Tamaki \textit{et al.} consider a protocol in which Alice prepares imperfect eigenstates of the Z basis, $\ket{\phi_{0Z}}$ and $\ket{\phi_{1Z}}$, and only one imperfect eigenstate of the X basis, $\ket{\phi_{0X}}$.
They consider single-photons, although the results can be generalized to phase-randomized weak coherent pulses using the decoy analysis \decoyrefs.
Following the notation already introduced in \cref{section:protocol}, we denote the probability that Alice chooses the Z (X) basis $p_Z^A$ ($p_X^A$).
The measurements done by Bob are described by the POVMs $\{M_{0\beta}, M_{1\beta}, M_f\}$, where $M_f$ corresponds to an inconclusive outcome and $M_{j\beta}$ corresponds to the outcome of bit value $j$ for the basis $\beta\in \{Z, X\}$.
His basis choice probabilities are denoted with $p_Z^B$ and $p_X^B$.
Note that, since $M_f$ is considered to be the same for both basis, it is assumed that the efficiencies of the detectors are the same.
To prove security, Tamaki \textit{et al.}  consider an equivalent scenario where Alice can prepare the entangled states given by
\begin{equation}
	\ket{\Psi_Z}_{AB} = \frac{1}{\sqrt{2}}\sum_{j=0,1}\ket{j_Z}_A\ket{\phi_{jz}}_{B}  \label{eq:Psi_Z}
\end{equation}
and
\begin{equation}
	\ket{\Psi_X}_{AB} = \ket{0_X}_A\ket{\phi_{0x}}_{B}. \label{eq:Psi_X}
\end{equation}
Then, she measures in the system $A$ in the Z(X) basis if she prepared $\ket{\Psi_Z}_{AB}$( $\ket{\Psi_X}_{AB}$). The outcome of the measurement is the bit value that Alice chooses in the real scenario. Finally, she sends system B to Bob. The phase error rate can be calculated as the fictitious QBER in the X basis when Alice prepared $\ket{\Psi_Z}_{AB}$, i.e.,
\begin{equation}
	Q_X = \frac{Y_{0x,1x}^{(Z)} + Y_{1x,0x}^{(Z)}}{Y_{0x,0x}^{(Z)}+Y_{0x,1x}^{(Z)}+Y_{1x,0x}^{(Z)}+Y_{1x,1x}^{(Z)}}. \label{eq:single-photon_phase_error_rate}
\end{equation}
Here  $Y_{jx,ix}^{(Z)}$ is the joint probability that Alice and Bob obtain the bits $i$ and $j$, respectively, when they measure in the X basis given the preparation of the Z state $\ket{\Psi_Z}_{AB}$.
In \cite{tamaki_loss-tolerant_2014}, it is shown that the three states alone are enough to calculate \cref{eq:single-photon_phase_error_rate}, as it is possible to estimate all the virtual yields from the ones that would be obtained in the real protocol.

In our protocol, the POVM elements of the monitoring line are given by the following \cite{rusca_security_2018}:
\begin{eqnarray}
    &M_{t0}=\frac{1}{4}\ket{0}\bra{0}= \frac{1}{4}M_{0Z}  \label{eq:M_t0}, \\
    &M_{t1}=\frac{1}{2}\ket{-}\bra{-}= \frac{1}{2}M_{1X} \label{eq:M_t1}, \\
    &M_{t2}=\frac{1}{4}\ket{1}\bra{1}= \frac{1}{4}M_{1Z} \label{eq:M_t2}, \\
    &M_{f}=\frac{1}{2}\mathcal{I} + \frac{1}{2}\ket{\emptyset}\bra{\emptyset} + \frac{1}{4}(\ket{0}\bra{1}+ \ket{1}\bra{0}).  \label{eq:M_f}
\end{eqnarray}
The detections in the middle time bin correspond to the projection to the state $\ket{-}$, but we do not have access to the projection to $\ket{+}$. To solve this, as was done in \cite{rusca_security_2018}, we use the fact that the side bins are projections to the Z states. 
This allows us to estimate the missing statistics from linear combinations of experimentally available ones, since
\begin{equation}
	M_{0X}=\ket{+}\bra{+} = \ket{0}\bra{0} + \ket{1}\bra{1}-\ket{-}\bra{-} = M_{0z}+M_{1z} - M_{1x}. \label{eq:M_0X}
\end{equation}
Using this, we are able to apply the LT security proof. For a more detailed explanation, we refer the reader to \cref{appendix:security proof}. Note that, assuming perfect state preparation, \textit{i.e.} $\ket{\phi_{0z}}=\ket{0}$, $\ket{\phi_{1z}}=\ket{1}$ and $\ket{\phi_{0x}}=\ket{+}$, our method yields the same result as in \cite{rusca_security_2018}. The reduction to the perfect state preparation case is shown in \cref{section:reduction perfect}.

The $Q_X$ obtained with \cref{eq:single-photon_phase_error_rate} is the phase error rate assuming single-photons in the asymptotic regime. 
Because we are using attenuated laser pulses instead of single-photons, we apply the one-decoy analysis in the finite-key regime \cite{rusca_finite-key_2018} to generalize the security to phase-randomized weak-coherent pulses. 
Then, the final phase error rate, which we denote with $\phi_Z$, is given by
\begin{equation}
	\phi_Z = \overline{Q}_X + \gamma(\epsilon_{sec}, \overline{Q}_X, \underline{s}_Z, \underline{s}(X_{side}, Z)), \label{eq:phase_error_rate_decoy}
\end{equation}
where 
\begin{equation}
	\gamma(a,b,c,d) = \sqrt{
	\frac{(c+d)(1-b)b}{cd\log 2}\log_2\left(\frac{(c+d)21^2}{cd(1-b)ba^2}\right)
	},
\end{equation}
$\epsilon_{sec}$ is the secrecy parameter, $\underline{s}_Z$ is a lower bound on the single-photon events in which Alice sent a Z state and Bob got a detection in the Z basis, $\underline{s}(X_{side}, Z)$ is a lower bound of the single-photon events in which Bob gets a detection in a side bin of the X basis and Alice sent a Z state and $\overline{Q}_X$ is an upper bound on $Q_X$.
For more details about the security analysis, we refer the reader to \cref{appendix:security proof}.

We remark that we are only considering collective attacks, but the security proof could be generalized to coherent attacks by using Azuma's inequality \cite{azuma_weighted_1967,boileau_unconditional_2005,tamaki_unconditional_2009} or the quantum de Finetti theorem\cite{caves_unknown_2002,konig_finetti_2005}.

\section{Secret key exchange}\label{section:secret-key-exchange}
We performed the secret key exchange over three different QC lengths, namely 101.4km, 112.6km and 151km. We employed ULL fiber with a loss of 0.17db/km. To estimate the phase error rate, we used the analysis of Rusca \textit{et al} and our adapted version of the LT method. 
For the LT calculation, we use the data presented in \cref{tab:table1} with the signal level, $\mu_0=0.5$, since the LT method does not consider correlations between intensities and qubit encoding.
Furthermore,  note that this analysis only requires the projection of the characterized states to the X-Z plane of the Bloch sphere. 
For this reason, $\theta$ has to be adjusted for the 0Z and 1Z states when they satisfy $\varphi>\pi/2$ so that the X component has the correct sign. 
This sign becomes relevant when calculating the virtual states in the security proof.

\begin{figure}[h!]
	\centering
	\includegraphics[width=0.65\linewidth]{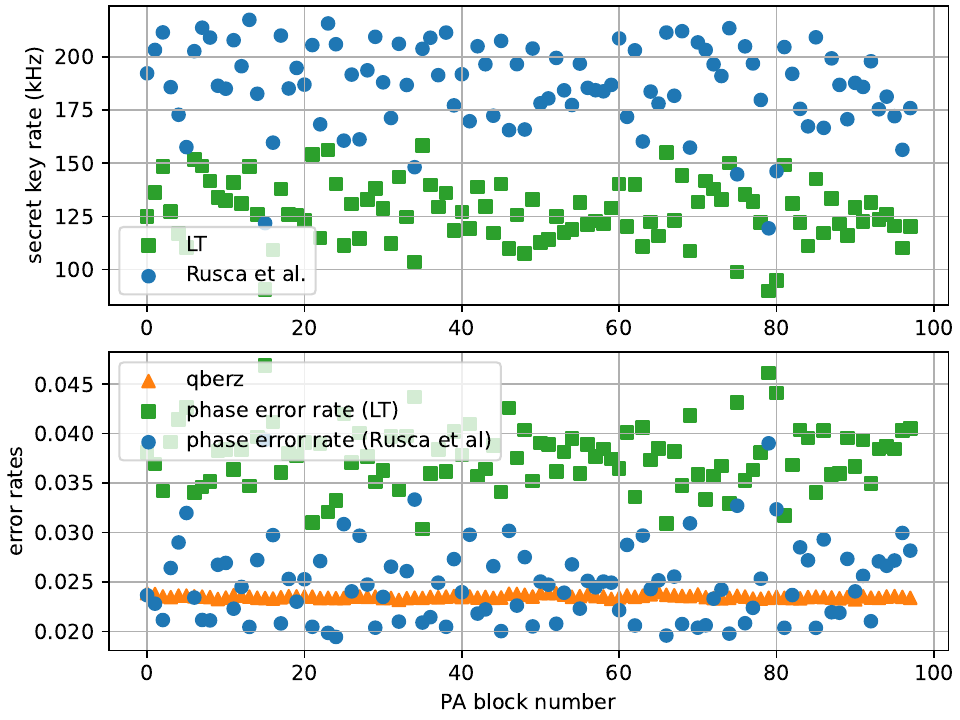}
	\caption{Performance with 101.4km. The curves denoted as Rusca \textit{et al}. use the results of \cite{rusca_security_2018}, whereas the ones denoted as LT use a combination of \cite{rusca_security_2018,tamaki_loss-tolerant_2014}. For the LT results we assume $\theta_{0z}=\rm 0.222 rad$, $\theta_{1z}=\rm 3.425 rad$ and $\theta_{0x}=\rm 1.66 rad$.}
	\label{fig:performance_LT_101km}
\end{figure}

The block size was set to $8\times10^6$. The mean photon numbers are $\mu_0=0.5$ and $\mu_1=0.23$. Alice's probabilities are changed for each distance to optimize the SKR. We consider secrecy and correctness parameters of $\epsilon_{sec}= 10^{-9}$ and $\epsilon_{ec}= 10^{-15}$, respectively. As mentioned in \cref{section:setup}, EC is not implemented but the leaked information is calculated assuming an LDPC algorithm is implemented. For a QBERZ of around 2\% and a syndrome size of 1/3, we consider that 33.3\% of the bits are leaked \cite{gruenenfelder_performance_2022,constantin_fpga-based_2017}. Similarly, PA is not implemented but the final SKR is calculated from the statistics and estimated phase error rate. \cref{fig:performance_LT_101km} shows the performance of our protocol with the analysis of \cite{rusca_finite-key_2018} and the LT analysis for a QC length of 101.4km as an example. The QBERZ at 101.4km is the highest because of a worse dispersion compensation. This is also the reason of the lower SKR compared to the QC length of 112.6km.

The results displayed in \cref{tab:table2} show that assuming perfect states leads to an overestimation of the SKR. 
Overall, the results highlight the importance of characterization of imperfections and its use in the security analysis.

\begin{table}[h!]
	\begin{center}
		\caption{Secret key exchange results. The results using the analysis from \cite{rusca_security_2018} are denoted as $\phi_Z$ and $SKR$ for the phase error rate and secret-key rate, respectively. On the other hand, the results using our analysis are marked with an $(LT)$ superindex. For this analysis, we use the $\theta$ values shown in \cref{tab:table1} for the signal intensity ($\mu_0$). The results shown in the table are calculated with the mean of all the PA blocks acquired during the measurement. 
        }
		\label{tab:table2}
		\begin{tabular}{ccccccccc}\hline
			QC length (km) & $p_Z^A$ (\%) & $p_{\mu0}$ (\%) & QBERZ (\%) & $\phi_Z$ (\%) & $SKR$ (bps) & $\phi_Z^{(LT)}$ (\%) & $SKR^{(LT)}$ (bps)  \\ \hline
			101.4 & 80.5 & 30.1 & $2.35$ & $2.49$ & $1.87\times10^5$ & $3.77$ & $1.27\times 10^{5}$ \\
			112.6 & 80.5 & 41.4 & $1.89$ & $3.03$ & $1.91\times 10^5$ & $3.52$ & $1.73\times 10^5$  \\
			151.0 & 76.6 & 47.3 & $1.73$ & $2.30$ & $3.04\times 10^4$ & $3.74$ & $1.85\times 10^4$  \\ \hline
		\end{tabular}
	\end{center}
\end{table}

\newpage

\section{Discussion}\label{section:discussion}
In this work, we present a simplified BB84 protocol with time-bin encoding and one decoy level. 
The performance of the system could be improved, but high secret-key rates have already been demonstrated in previous works \cite{grunenfelder_fast_2023} and over a QC length of 421km \cite{boaron_secure_2018} using the same protocol.
Instead, our focus is on improving the security of the protocol.
We have demonstrated a simple characterization method that uses Bob to analyze the states and does not require any changes in the setup.
This allowed us to measure the SPFs instead of merely relying on low QBERs.
Moreover, we have adapted the LT method to our protocol so that the estimated SPFs can be taken into account for the phase error rate estimation.

Through these methods, we report several relevant findings. 
Firstly, the state characterization reveals correlations between intensity levels and bit-and-basis encoding. 
This is caused by the IM implementing both the intensity levels and the qubit encoding simultaneously.
This effect could be mitigated by adding another IM so that the processes are independent.
However, because of the high repetition rate we would still expect the IMs to also incorporate intensity and bit-and-basis correlations \cite{trefilov_intensity_2025,agulleiro_modeling_2025}.
Although these side-channels have been addressed independently by security proofs \cite{xing_cross_correlations_2025,pereira_quantum_2020,curras-lorenzo_security_2023,sixto_security_2022}, there is still an urgent need to be able to combine all of them in one.

Secondly, when including the SPFs in the parameter estimation, we conclude that making the invalid assumption that the state preparation is perfect may lead to underestimating the phase error rate.
This result emphasizes that the push for higher key rates should also be accompanied by a greater care about the imperfections that high speed operation causes and its effect on the security of the protocol.
Still, we show that it is possible to securely distribute keys at a rate of around 18.5kbps over a distance of 151km even in the presence of SPFs, which represents a reduction of less than 40\% compared to the SKR obtained without consideration of imperfections.

The security proof only considers collective attacks and, like the preceding one \cite{rusca_security_2018}, is still missing the squashing map. The next steps are to address these issues, as well as to characterize known side-channels using techniques like the ones presented in \cite{trefilov_intensity_2025,wiesemann_evaluation_2025,agulleiro_modeling_2025,xing_cross_correlations_2025,xing_characterization_2024,marcomini_characterising_2025,marcomini_loss-tolerant_2025}, and to incorporate the results into the security analysis.

\section{Acknowledgments}

We acknowledge support from the Galician Regional Government (consolidation of Research Units: AtlantTIC and own funding through the ``Planes Complementarios de I+D+I con las Comunidades Autónomas'' in Quantum Communication), MICIN with funding from the European Union NextGenerationEU (PRTR-C17.I1), the ``Hub Nacional de Excelencia en Comunicaciones Cu{\' a}nticas'' funded by the Spanish Ministry for Digital Transformation and the Public Service and the European Union NextGenerationEU, the European Union's Horizon Europe Framework Programme under Marie Sk\l{}odowska-Curie grant 101072637 (project QSI) and project ``Quantum Security Networks Partnership'' (QSNP; grant 101114043), and the European Union via the European Health and Digital Executive Agency (HADEA) under project QuTechSpace (grant 101135225).
A. A. acknowledges support through the ``Programa de axudas á etapa predoutoral da Xunta de Galicia (Consellería de Educación, Ciencia, Universidades e Formación Profesional)'' cofunded by the European Union in the frame of the program FSE+ Galicia 2021-2027.
D. R. acknowledges the Ramón y Cajal Grant RYC2024-050943-I, funded by MICIU/AEI/10.13039/501100011033 and by the European Social Fund Plus (ESF+).


\clearpage

\appendix

\newpage
\section{Impact of delay mismatch on the visibility}\label{appendix:delay_mismatch}
The performance of our protocol depends greatly on the quality and similarity of Alice's and Bob's interferometers.
If the arms of the interferometers have a big imbalance of intensity, this will affect the quality of interference.
On the other hand, if their delays are significantly different relative to the FWHM of the laser pulses used, the interference is also expected to be worse.
The latter effect is more detrimental when the laser pulses have chirp.
All of these effects have been thoroughly studied by Millet \textit{et al.} for time-bin encoding protocols like ours \cite{millet_influence_2025}.

\begin{figure}[h!]
	\centering
	\includegraphics[width=0.5\linewidth]{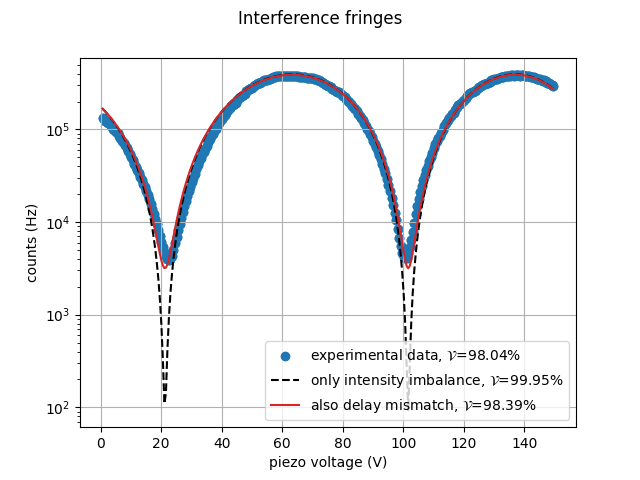}
	\caption{Counts plotted in logarithmic scale against the applied voltage on the piezo. The blue points correspond to the experimental data.
    The dashed black line corresponds to using \cref{eq:naive_fringes}.
    The solid red line corresponds to using \cref{eq:tau_fringes} with $\mathcal{V}_{\Delta \tau}=0.984$.
    For both curves, the measured $n(X0)$ and $n(X2)$ were used.}
	\label{fig:visibility}
\end{figure}

The measured delay mismatch of our interferometers is of 4ps, which is almost 11\% of the FWHM of the laser pulses. 
This is expected to have a noticeable effect on the visibility according to the results of Millet \textit{et al}.
In our setup, we quantify these imperfections in the following way.
As in the secret key exchange, we gain-switch the laser and create the coherent pairs by letting the pulses go through Alice's interferometer.
However, we do not modulate the intensity. 
Then, we count the detections on Bob's early, middle and late time bins while we perform a sweep of the voltage applied to the piezoelectric component of Alice.
The voltage sweep changes the relative phase between the interferometers, which causes interference fringes to appear in the middle time bin (see blue points in \cref{fig:visibility}).
As a figure of merit, we estimate the visibility, $\mathcal{V}$, given by
\begin{equation}
    \mathcal{V}=\frac{\max\{n(X1)\} - \min\{n(X1)\}}{\max\{n(X1)\} + \min\{n(X1)\}}.
\end{equation}
For this measurement, we connect Alice and Bob with single-mode fiber and a variable attenuator instead of ULL fiber.

An imbalance of the losses of the arms of the interferometers will cause the counts on the side bins, $n(X0)$ and $n(X2)$, to be different.
Moreover, this will also cause the middle bin to have a visibility below 1.
This can be easily understood by writing down the counts in the middle bin, $n(X1)$, as a function of the side bins and the phase, $\varphi$, that is
\begin{equation}
    n(X1)=n(X0) + n(X2) - 2\sqrt{n(X0)n(X2)}\cos\varphi. \label{eq:naive_fringes}
\end{equation}
Then, if $n(X0)=n(X2)$, we get a visibility of 1.
On the contrary, $n(X0)\neq n(X2)$ implies $\mathcal{V}<1$.

The dashed black line in \cref{fig:visibility} was calculated with \cref{eq:naive_fringes} using the measured $n(X0)$ and $n(X2)$. 
The significant difference with the experimental data indicates that the intensity imbalance alone does not explain the visibility that we measure.
This is because other factors, like the delay mismatch and chirp of the pulses, come into play \cite{millet_influence_2025}.
We model the other imperfections by adding a factor in front of the cosine function.
We call this factor $\mathcal{V}_{\Delta \tau}$, as it is the expected visibility if there was no intensity imbalance but there is a delay mismatch. Then,
\begin{equation}
    n(X1)=n(X0) + n(X2) - 2\mathcal{V}_{\Delta \tau}\sqrt{n(X0)n(X2)}\cos\varphi. \label{eq:tau_fringes}
\end{equation}
The solid red line in \cref{fig:visibility} is plotted by optimizing $\mathcal{V}_{\Delta \tau}$ to fit the experimental data with \cref{eq:tau_fringes}, while $n(X0)$ and $n(X2)$ are still fixed to the measured values. With the fit, we obtain a value of $\mathcal{V}_{\Delta \tau}=0.984$.

\newpage
\section{State characterization}\label{appendix:state characterization}
\subsection{Calculations}\label{appendix:state characterization calculations}

Let us denote Alice's chosen bit and basis with $A\in\{0Z,1Z,0X\}$. Let us also define the mean photon number setting as $m\in\{\mu_0,\mu_1\}$. Then, her prepared states may be written as 
\begin{equation}
    \ket{\psi_{A,m}} = \ket{\sqrt{\mu_{m, A}}\cos\frac{\theta_{A, m}}{2}}_{t0} \ket{e^{i\varphi_{A, m}}\sqrt{\mu_{m, A}}\sin\frac{\theta_{A, m}}{2}}_{t1},
\end{equation}
where $\theta_{A,m}$ and $\varphi_{A, m}$ are the polar and azimuthal angle in the Bloch sphere. 
In an ideal scenario, the bit and basis encoding and the decoy intensity encoding are independent and perfect, meaning that $\theta_{A,m}=\theta_A\in\{0, \pi, \pi/2\}$, $\varphi_{A,m}=\varphi_0=0$ and $\mu_{m,A}=\mu_m\in\{\mu_0,\mu_1\}$.
Thus, the coordinates in the Bloch sphere depend only on the bit and basis choice and $\mu$ depends only on the intensity setting. 
However, in practice and as shown in \cref{section:state-characterization}, the intensity modulation is not perfect, which introduces SPFs. 
In our specific scenario, this also introduces correlations between the bit and basis encoding and the intensity levels since they are both implemented by the same IM. 
For this reason, $\mu_{m,A} \in\{\mu_0 + \delta \mu_{0,A}, \mu_1 + \delta \mu_{1,A}\}$ is allowed to depend on the bit-and-basis choice and the angles $\theta_{A,m} \in \{\delta \theta_{0Z, m}, \pi + \delta \theta_{1Z, m}, \pi/2 + \delta\theta_{0X, m}\}$ and $\varphi_{A, m} \in \{\delta\varphi_{0Z, m}, \delta\varphi_{1Z,m}, \delta\varphi_{0X,m}\}$ are allowed to depend on the intensity setting.

With these considerations, the number of events in which Alice prepares state $A$ with intensity setting $m$ and Bob gets a detection in the early and late time bins of the Z basis are given by
\begin{equation}
	n(m, Z0, A)  = C_Z(m, A) {\cos}^2 (\theta_{A,m}/2)
\end{equation}
and
\begin{equation}
	n(m, Z1, A)  = C_Z(m, A) {\sin}^2 (\theta_{A,m}/2) ,
\end{equation}
respectively. 
Here, $C_Z(m, A)$ is defined as
\begin{equation}
	C_Z(m, A) \equiv Np_A^A p_{m}\mu_{m,A} \eta_Zp_Z^B \eta_{QC},
\end{equation}
where $N$ is the total number of counts, $p_A^A$ is the probability that Alice sends the state $A$, $p_{m}$ is the probability that she chooses the intensity setting $m$, $\eta_Z$ is the efficiency of the detector in the Z basis, $p_Z^B$ is the probability that the state is measured in the Z basis and $\eta_{QC}$ is the transmission of the quantum channel.
Using that $C_Z(m, A)=n(m, Z0, A)+n(m, Z1, A)$, one can calculate $\theta$ as
\begin{equation}
	\theta_{A, m} = 2 \arccos\sqrt\frac{n(m, Z0, A)}{n(m, Z0, A)+n(m, Z1, A)}.
\end{equation}

The detections in the early, middle and late time bins of the X basis are given by
\begin{equation}
	n(m, X0, A)  = C_X(m, A) {\cos}^2 (\theta_{A,m,X}/2), \label{eq:counts_X0}
\end{equation}
\begin{equation}
	n(m, X1, A)  = C_X(m, A) (1-2\sin(\theta_{A,m,X}/2)\cos(\theta_{A,m,X}/2)\mathcal{V}_{\Delta\tau}\cos\varphi_{A, m}) \label{eq:counts_X1}
\end{equation}
and
\begin{equation}
	n(m, X2, A)  = C_X(m, A) {\sin}^2 (\theta_{A,m,X}/2), \label{eq:counts_X2}
\end{equation}
respectively.
The motivation of adding the factor $\mathcal{V}_{\Delta \tau}$ in \cref{eq:counts_X1} is explained in \cref{appendix:delay_mismatch}.
Here, we define 
\begin{equation}
	C_X(m, A) \equiv Np_A^A p_{m}\mu_{m,A} \frac{1}{4}\eta_Xp_X^B \eta_{QC},
\end{equation}
where $\eta_X$ is the efficiency of the detector in the X basis and $p_X^B$ is the probability that Bob measures in the X basis.
Similarly to $C_Z$, $C_X$ can be calculated by the sum of the counts in the side bins.
Thus, $\theta_{A, m}$  can be calculated again in a similar fashion as
\begin{equation}
	\theta_{A, m} = 2 \arccos\sqrt\frac{n(m, X0, A)}{n(m, X0, A)+n(m, X2, A)}.
\end{equation}
In order to estimate $\varphi_{A, m}$, we use the middle bin with the formula given by
\begin{equation}
	\varphi_{A, m} = \arccos\left( \frac{n(m, X0, A) + n(m, X2, A) - n(m, X1, A)}{2\mathcal{V}_{\Delta \tau}\sqrt{n(m, X0, A)n(m, X2, A)}} \right).
\end{equation}

\begin{figure}[h!]
	\centering
	\includegraphics[width=0.75\linewidth]{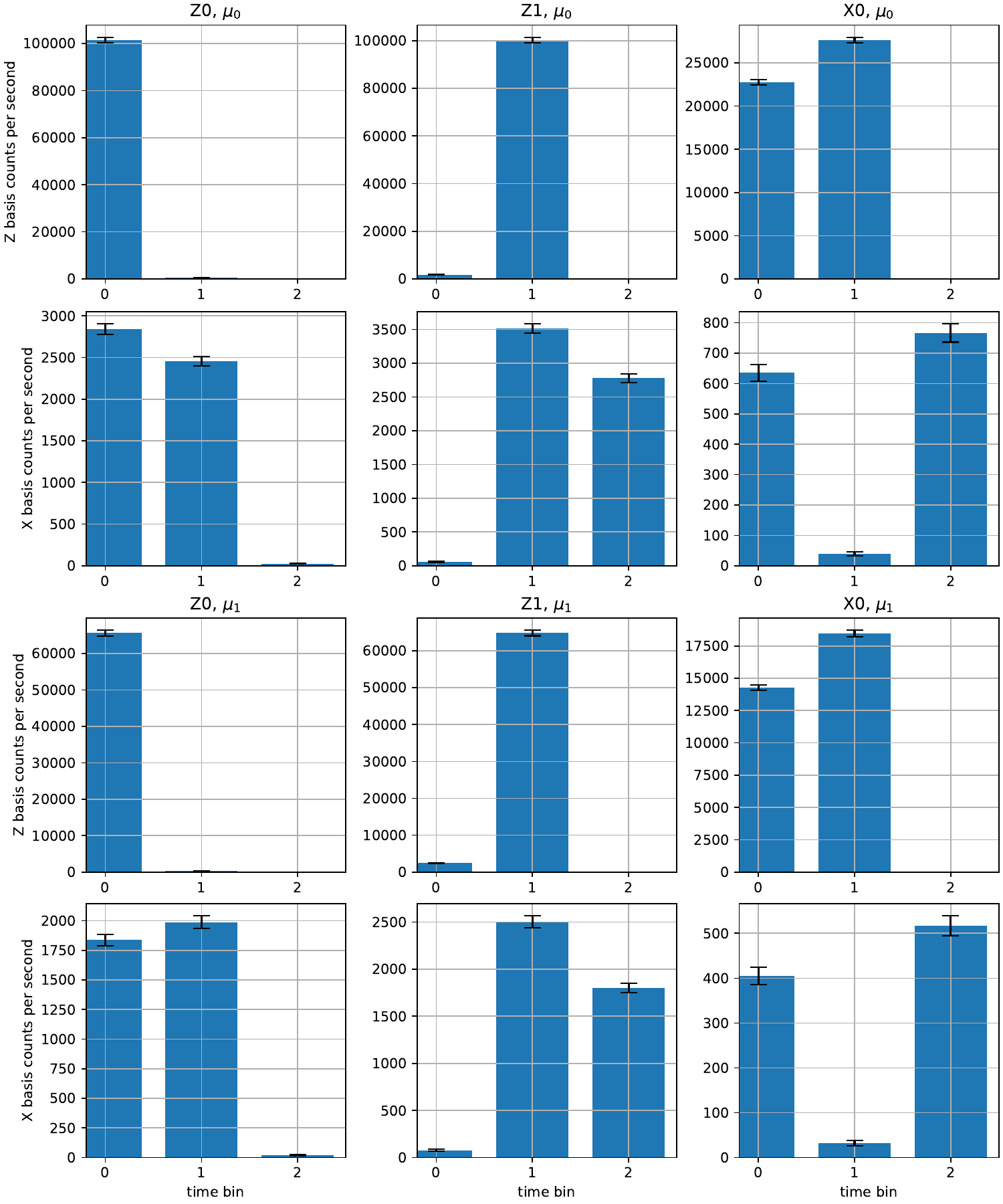}
	\caption{Logged statistics with 25 dB attenuation. The error bars are calculated with the standard deviation.}
	\label{fig:statistics_101km}
\end{figure}

Note that for the previous calculations, we assume that the dark counts are negligible relative to the actual counts due to the states.
\cref{fig:statistics_101km} shows the histograms used for the state characterization, whose results are displayed in \cref{tab:table1}.

From the parameters $C_{Z,X}(m, A)$ one can also calculate ratios between decoy levels.
The ratio between the intensities resulting of the same intensity setting $m$ but different encoding, $A$ and $A'$, can be calculated as
\begin{equation}
	\frac{\mu_{m,A}}{\mu_{m,A'}} = \frac{C_X(m, A)}{C_X(m, A')}\frac{p_{A'}^A}{p_A^A}. \label{eq:ratio_mus_1}
\end{equation}
Similarly, the ratio between the intensities when choosing different intensity levels, $m$ and $m'$, but using the same encoding $A$ is given by
\begin{equation}
	\frac{\mu_{m,A}}{\mu_{m',A}} = \frac{C_X(m, A)}{C_X(m', A)}\frac{p_{m'}}{p_{m}}. \label{eq:ratio_mus_2}
\end{equation}
The same could be calculated with data from the Z basis. \cref{fig:intensities_101km} was plotted using \cref{eq:ratio_mus_1,eq:ratio_mus_2}.

\subsection{Characterization schemes}\label{appendix:characterization configurations}

\begin{figure}[h!]
	\centering
	\includegraphics[width=0.75\linewidth]{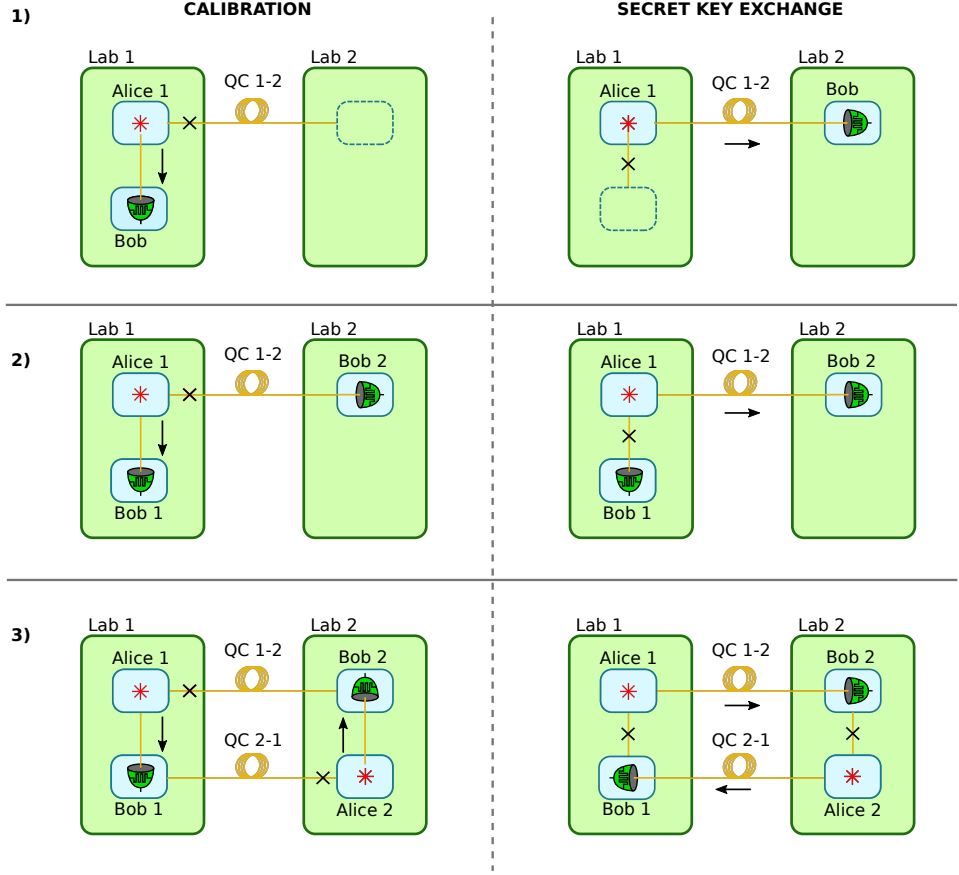}
	\caption{Drawing of different schemes for state characterization. The green boxes represent trusted areas like Alice's and Bob's labs. We assume Eve has no access to these areas. Black crosses indicate the fiber is not connected. (1) Alice 1 and Bob perform the state characterization in the same lab. For the secret key exchange, Bob is transported to lab 2, where he can connect to the untrusted QC. (2) Alice 1 has a replica of Bob, denoted Bob 1, to whom she sends the states to perform state characterization. To perform secret key exchange, she sends the states to Bob 2 in lab 2. (3) Alice 1 (2) and Bob 1 (2) perform state characterization in lab 1 (2). Secret key exchange is performed between Alice 1 (2) and Bob 2 (1) in two directions between the two labs. }
	\label{fig:calibration_schemes}
\end{figure}

One could use different schemes to perform the state characterization, as depicted in  \cref{fig:calibration_schemes}. We assume that it is all performed in a trusted area to which Eve has no access. For example, after building the transmitter and receiver in the same lab, one could connect them and directly characterize the SPFs. A VOA may be used such that the detectors are not saturated. Then, Bob is sent to the other lab and both Alice and Bob connect themselves to the untrusted QC to perform the secret key exchange. 
In this scenario, one would have to assume that the SPFs stay constant, as transporting Bob back and forth between the two locations is too unpractical.
Alternatively, one could place a replica of Bob next to Alice. 
Then, Alice would use the replica of Bob for calibration and the real Bob for the secret key exchange.
Even further, at each location, one could have an Alice and a Bob. 
This would allow them to perform the calibration in the trusted areas, and to perform the key exchange in both directions between the two locations. 
The advantage of the first one is that it is cheaper, while the other two would allow to periodically switch between calibration and key exchange, thus monitoring any possible changes.

\clearpage
\section{Applying the loss-tolerant method to our protocol}\label{appendix:security proof}

In the three-state BB84 protocol with time-bin encoding and a simplified measurement scheme \cite{rusca_security_2018}, Alice sends only three states ($\ket{0}$, $\ket{1}$, $\ket{+}$) and Bob can project to $\ket{0}$, $\ket{1}$ and only one eigenstate of the X basis, namely $\ket{-}$. The security proof described by Rusca \textit{et al.} only considers perfect state preparation \cite{rusca_security_2018}. On the other hand, the LT method \cite{tamaki_loss-tolerant_2014} includes SPFs, but considers that Bob can also project to the $\ket{+}$ state. Here, we combine both security proofs to achieve security for the simplified BB84 protocol in the presence of SPFs.

In \cref{section:loss-tolerant}, a summary of the original LT method is presented to show how it works and introduce the notation that will be followed in the rest of this Appendix. The new contributions to adapt the loss-tolerant method to the simplified measurement scheme are presented in \cref{section:adaptation}. Note that, following the reasoning of Tamaki \textit{et al.}, we assume that the states lie on the X-Z plane of the Bloch sphere, i.e., $\varphi=0$. However, the results are general for any $\varphi$ as long as the three states are mutually linearly independent \cite{tamaki_loss-tolerant_2014}.

\subsection{Original loss-tolerant method}\label{section:loss-tolerant}

Tamaki \textit{et al.} consider a three-state BB84 protocol described by the following steps:
\begin{itemize}
	\item Alice prepares imperfect states $\ket{\phi_{j\beta}}$ with $j\beta \in \{0Z,1Z, 0X\}$. Alice chooses the Z basis with probability $p_Z^A$. The Bloch vectors of the states are given by $(P_X^{j\beta}, P_Y^{j\beta}, P_Z^{j\beta})$.
	\item Bob chooses his measurement basis with probabilities $p_Z^B$ and $p_X^B=1-p_Z^B$. His POVMs are denoted as $\{M_{0\beta}, M_{1\beta}, M_f\}$ with $\beta \in\{X,Z\}$ and where $M_f$ corresponds to an inconclusive outcome, and $M_{j\beta}$ corresponds to the outcome of bit value j.
	\item Bob reveals the detections in the X basis for parameter estimation.
\end{itemize}
As briefly introduced in \cref{section:security proof}, in \cite{tamaki_loss-tolerant_2014}, security is proved by considering an equivalent virtual scenario where Alice prepares entangled states $\ket{\Psi_Z}_{AB}$ and $\ket{\Psi_X}_{AB}$ given by \cref{eq:Psi_Z,eq:Psi_X} and calculating the phase error rate $Q_X$ expressed in \cref{eq:single-photon_phase_error_rate}.
Let us denote with $Y_{s\beta,j\alpha}^{(Z)}$ ($Y_{s\beta,j\alpha}^{(X)}$) the probability that Alice prepares $\ket{\Psi_Z}_{AB}$ ($\ket{\Psi_X}_{AB}$), Alice measures $j\alpha\in\{0Z, 1Z, 0Z, 1Z\}$ and Bob measures $s\beta\in\{0Z, 1Z, 0Z, 1Z\}$. 
Note that the yields $Y_{sx,0z}^{(Z)}$, $ Y_{sx,1z}^{(Z)}$ and $Y_{sx,0x}^{(X)}$ of the virtual protocol with $s\in\{0,1\}$ are equivalent to the yields of the real protocol.
The virtual yields needed for the calculation of the phase error rate with \cref{eq:single-photon_phase_error_rate}, on the other hand, can be calculated as
\begin{equation}
	Y_{sx,jx}^{(Z)} = Tr\left[\sigma_{B;jx, vir}\right]
	Tr\left[D_{sx}\sigma_{B;jx, vir}^{\prime}\right]p_X^{B}, \label{eq:virtual_yields_LT}
\end{equation}
where $Tr\left[\sigma_{B;jx, vir}\right]$ is the probability that Alice emits the virtual state 
\begin{equation}
	\sigma_{B;jx,vir} = Tr_{A}\left[ \ket{j_X}\bra{j_X}_A \otimes
	\mathcal{I}_{B}\ket{\Psi_Z}\bra{\Psi_Z}_{AB}  
	\right]. \label{eq:4}
\end{equation}
Here, $\sigma_{B;jx, vir}^{\prime}$ is the normalized virtual state. 
One can consider them to lie on the X-Z plane of the Bloch sphere, as this assumption can be lifted by using a filtering operation and the security still holds \cite{tamaki_loss-tolerant_2014}. 
Thus, the normalized state may be written as $\sigma_{B;jx, vir}^{\prime} = \frac{1}{2}(\mathcal{I} + P_X^{jx,vir}\sigma_X + P_Z^{jx,vir}\sigma_Z)$, where $\mathcal{I}$ is the identity, $P_X^{jx,vir}$ ($P_Z^{jx,vir}$) is the X (Z) component of the Bloch vector of $\sigma_{B;jx, vir}^{\prime}$, and  $\sigma_X$ and $\sigma_Z$ are the X and Z Pauli matrices, respectively. 
Finally, $D_{sx}=\sum_kA_k^{\dagger}M_{sx}A_k$,with $A_k$ being an arbitrary operator representing Eve's action.  

The Bloch vectors of the virtual states and their probabilities to be emitted can be calculated from the characterization of the real states (see \cref{section:bloch vectors}). Therefore, the task of estimating the virtual yields reduces to calculating the transmission rates given by
\begin{equation}
	q_{sx|t} =  Tr\left[D_{sx}\sigma_t\right]/2, \label{eq:6}
\end{equation}
with $s\in\{0,1\}$ and $t\in\{Id, x, z\}$. 
This can be done using the experimentally accessible yields with the following system of equations:
\begin{eqnarray}
    Y_{sx,0z}^{(Z)}=\frac{1}{2}p_Z^Ap_X^B Tr\left[D_{sx}(\mathcal{I}+P_X^{0Z}\sigma_x + P_Z^{0Z}\sigma_Z)/2\right] =  \frac{1}{2}p_Z^Ap_X^B(q_{sx|Id} + P_Z^{0Z}q_{sx|z} + P_X^{0Z}q_{sx|x}), \label{eq:7}\\
    Y_{sx,1z}^{(Z)} =\frac{1}{2}p_Z^Ap_X^B Tr\left[D_{sx}(\mathcal{I}+P_X^{1Z}\sigma_x + P_Z^{1Z}\sigma_Z)/2\right] =  \frac{1}{2}p_Z^Ap_X^B(q_{sx|Id} + P_Z^{1Z}q_{sx|z} + P_X^{1Z}q_{sx|x}), \label{eq:8} \\
    Y_{sx,0x}^{(X)} =p_X^Ap_X^B Tr\left[D_{sx}(\mathcal{I}+P_X^{0X}\sigma_x + P_Z^{0X}\sigma_Z)/2\right] =(1-p_Z^A)p_X^B (q_{sx|Id} + P_Z^{0X}q_{sx|z} + P_X^{0X}q_{sx|x}). \label{eq:9}
\end{eqnarray}
Then, the obtained $q_{sx|t}$ are used in \cref{eq:virtual_yields_LT} and $Q_X$ can be calculated.

\subsection{Adaptation to our protocol}\label{section:adaptation}
Bob's POVM elements of the X basis in our protocol are given by \cref{eq:M_t0,eq:M_t1,eq:M_t2,eq:M_f}.
We denote the experimental yields as $Y_{ti, j\beta}^{(exp)}$ with $i\in \{0,1,2\}$ and $j\beta \in \{0Z, 1Z, 0X\}$, which is the joint probability of Bob getting a detection in time bin $t_i$ in the X basis and Alice sending state $j\beta$. Since $M_{t1}$ is the projection to $\ket{-}$ (except for a 1/2 factor), we can directly get the statistics with $s=1$ from the experimental yields as
\begin{equation}
	Y_{1x,j\beta}^{(Z)} = 2 Y_{t1,j\beta}^{(exp)}. \label{eq:21}
\end{equation}

Let us write down the other statistics that we have access to:
\begin{eqnarray}
    Y_{t0,jz}^{(exp)}=\frac{1}{4}Y_{0z,jz}^{(Z)}=\frac{1}{4}\frac{1}{2}p_Z^Ap_X^B(q_{0z|Id} + P_X^{jZ}q_{0z|x} + P_Z^{jZ}q_{0z|z}), \label{eq:24}\\
    Y_{t0,0x}^{(exp)}=\frac{1}{4}Y_{0z,0x}^{(X)}=\frac{1}{4}(1-p_Z^A)p_X^B(q_{0z|Id} + P_X^{0X}q_{0z|x} + P_Z^{0X}q_{0z|z}), \label{eq:25}\\
    Y_{t2,jz}^{(exp)}=\frac{1}{4}Y_{1z,jz}^{(Z)}=\frac{1}{4}\frac{1}{2}p_Z^Ap_X^B(q_{1z|Id} + P_X^{jZ}q_{1z|x} + P_Z^{jZ}q_{1z|z}), \label{eq:26}\\
    Y_{t2,0x}^{(exp)}=\frac{1}{4}Y_{1z,0x}^{(X)}=\frac{1}{4}(1-p_Z^A)p_X^B(q_{1z|Id} + P_X^{0X}q_{1z|x} + P_Z^{0X}q_{1z|z}). \label{eq:27}
\end{eqnarray}
Solving the system of equations given by \cref{eq:21,eq:24,eq:25,eq:26,eq:27}, one can calculate $q_{1x|t}$, $q_{0z|t}$ and $q_{1z|t}$ for $t \in \{Id, x, z\}$.
Then, using that $M_{0X}$ is a linear combination of $M_{0Z}$, $M_{1Z}$ and $M_{1X}$ as expressed by \cref{eq:M_0X}, one can calculate $q_{0x|t}$ as
\begin{equation}
	q_{0x|t} = q_{0z|t} + q_{1z|t} - q_{1x|t}. \label{eq:29}
\end{equation}
Thus, we have all the ingredients to calculate the phase error rate with  \cref{eq:single-photon_phase_error_rate}.

\subsubsection{Step-by-step solution}

\begin{enumerate}
\item With \cref{eq:21} solve for $q_{1x|t}$ with $t \in \{Id, x, z\}$. To do so, let us define the vectors
\begin{eqnarray}
    \mathbf{b_1} = \begin{pmatrix}
		Y^{(exp)}_{t1,0z}\\
		Y^{(exp)}_{t1,1z}\\
		Y^{(exp)}_{t1,0x}
	\end{pmatrix} 
    \quad \text{and} \quad
    \mathbf{q_s} = \begin{pmatrix}
		q_{sx|Id}\\
		 q_{sx|x}\\ 
		 q_{sx|z}
	\end{pmatrix} \text{ with } s\in\{0,1\}, \label{eq:definition_b1_and_qs}
\end{eqnarray}
and the matrix
\begin{equation}
B = {p_X^B}/{2} \begin{pmatrix}
    p_Z^A/2 & p_Z^AP_X^{0Z}/2 & p_Z^AP_Z^{0Z}/2\\
    p_Z^A/2 & p_Z^AP_X^{1Z}/2 & p_Z^AP_Z^{1Z}/2\\
    1-p_Z^A & (1-p_Z^A)P_X^{0X} & (1-p_Z^A)P_Z^{0X}\\
\end{pmatrix}.
\end{equation}
Note that the components of the Bloch vectors here should be known from the state characterization, as will be shown in \cref{section:bloch vectors}.
Then, $\mathbf{q_1}$ can be calculated as
 \begin{equation}\label{eq:solution_q1}
 	\mathbf{q_1} =B^{-1}\mathbf{b_1}.
 \end{equation}

\item With \cref{eq:21,eq:24,eq:25,eq:26,eq:27,eq:29}, solve for $q_{0x|t}$  with $t \in \{Id, x, z\}$. This can be done with the system of equations given by
\begin{equation}
    \mathbf{b_0} = B \mathbf{q_0},
\end{equation}
where we have defined
\begin{equation}\label{eq:definition_b0}
	\mathbf{b_0} = \begin{pmatrix}
		2Y^{(exp)}_{t0,0z}+2Y^{(exp)}_{t2,0z}-Y^{(exp)}_{t1,0z}\\
		2Y^{(exp)}_{t0,1z}+2Y^{(exp)}_{t2,1z}-Y^{(exp)}_{t1,1z}\\
		2Y^{(exp)}_{t0,0x}+2Y^{(exp)}_{t2,0x}-Y^{(exp)}_{t1,0x}
	\end{pmatrix}.
\end{equation} 
Then, the solution may be written as $\mathbf{q_0} =B^{-1}\mathbf{b_0}$, or, more generally, 
\begin{equation}
	\mathbf{q_s} =B^{-1}\mathbf{b_s} \quad \text{for} \quad s \in \{0,1\}.  \label{eq:solution_qs}
\end{equation}

\item Calculate $Y_{sx,jx}^{(Z)}$ using \cref{eq:virtual_yields_LT} with the obtained $q_{sx|t}$. 
\begin{eqnarray}
    Y_{0x,0x}^{(Z)} = p_X^{B} Tr\left[\sigma_{B;0x, vir}\right]\left(
	q_{0x|Id} + P_X^{0x, vir}q_{0x|x} + P_Z^{0x, vir}q_{0x|z}
	\right), \label{eq:36} \\
    Y_{1x,0x}^{(Z)} = p_X^{B} Tr\left[\sigma_{B;0x, vir}\right]\left(
	q_{1x|Id} + P_X^{0x, vir}q_{1x|x} + P_Z^{0x, vir}q_{1x|z}
	\right), \label{eq:37} \\
    Y_{0x,1x}^{(Z)} = p_X^{B} Tr\left[\sigma_{B;1x, vir}\right]\left(
	q_{0x|Id} + P_X^{1x, vir}q_{0x|x} + P_Z^{1x, vir}q_{0x|z}
	\right), \label{eq:38} \\
    Y_{1x,1x}^{(Z)} = p_X^{B} Tr\left[\sigma_{B;1x, vir}\right]\left(
	q_{1x|Id} + P_X^{1x, vir}q_{1x|x} + P_Z^{1x, vir}q_{1x|z}
	\right). \label{eq:39}
\end{eqnarray}
The traces and Bloch vectors of the virtual states are explicitly calculated in \cref{section:bloch vectors} from the state characterization.
We define the vectors 
\begin{equation}
	\mathbf{Y_{sx}^{(vir)}} = \begin{pmatrix}
		Y_{sx,0x}^{(Z)}\\
		Y_{sx,1x}^{(Z)}
	\end{pmatrix},
\end{equation}
with $s\in\{0,1\}$, and the matrix
\begin{equation}
	A = p_X^B \begin{pmatrix}
		Tr\left[\sigma_{B;0x, vir}\right] &  P_X^{0x, vir}Tr\left[\sigma_{B;0x, vir}\right] & P_Z^{0x, vir}Tr\left[\sigma_{B;0x, vir}\right]\\
		Tr\left[\sigma_{B;1x, vir}\right] &  P_X^{1x, vir}Tr\left[\sigma_{B;1x, vir}\right] & P_Z^{1x, vir}Tr\left[\sigma_{B;1x, vir}\right]
	\end{pmatrix}.
\end{equation}
Then, the virtual yields may be expressed in terms of the experimental yields as
\begin{equation}\label{eq:solution_yields}
	\mathbf{Y_{sx}^{(vir)}} = AB^{-1}\mathbf{b_s}.
\end{equation}
Note that $A$ and $B^{-1}$ have a factor of $p_X^B$ and $1/p_X^B$, respectively, such that they cancel out in the virtual yields. Thus, the result is independent of Bob's basis choice probabilities.

\item Calculate the phase error rate from the found virtual yields using \cref{eq:single-photon_phase_error_rate}.

\end{enumerate}

\begin{figure}[h!]
	\centering
	\includegraphics[width=\linewidth]{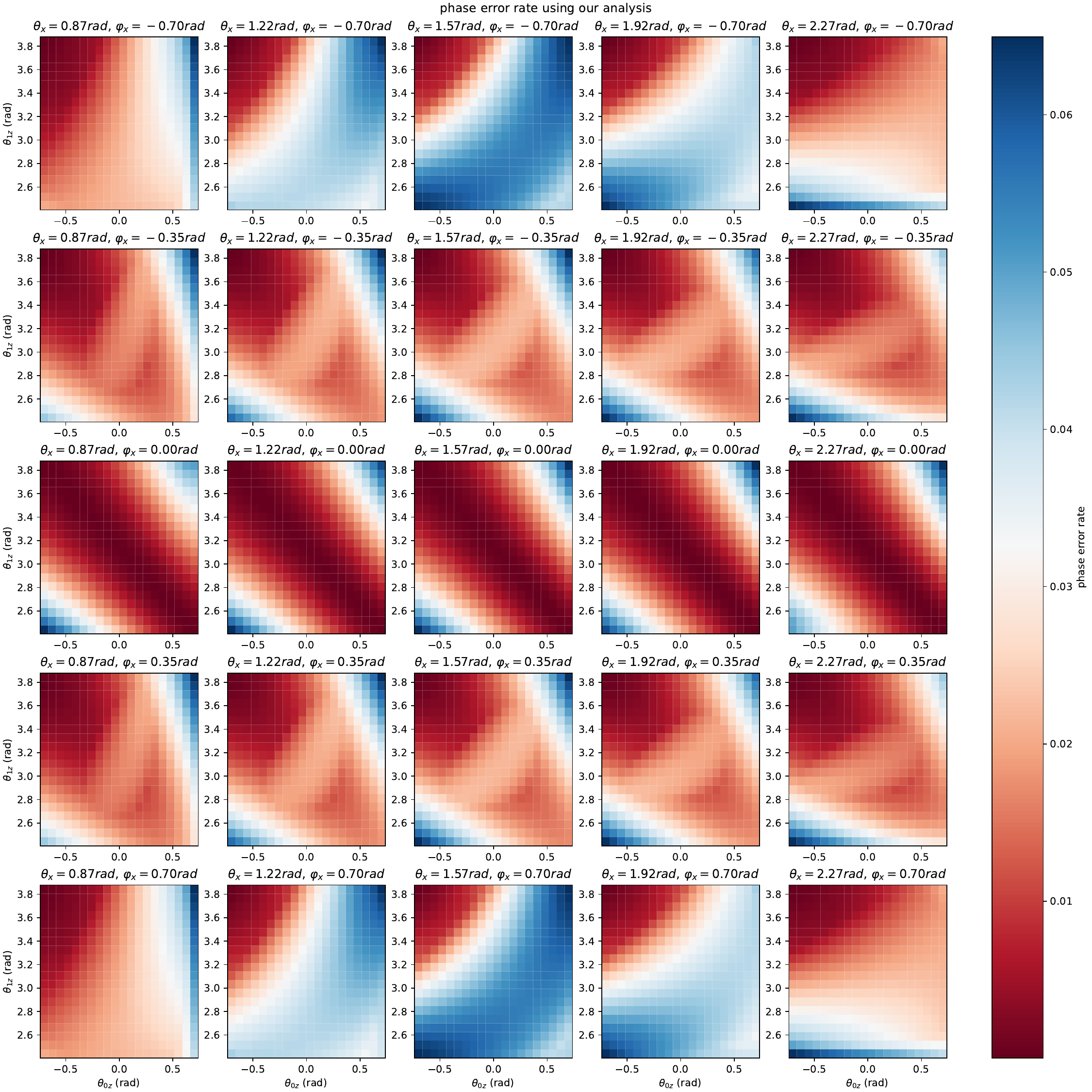}
	\caption{Simulation results using our analysis and considering single-photons. The probabilities of basis choice are fixed. The color indicates the phase error rate, while the X (Y) coordinates indicate $\theta_{0z(1z)}$.}
	\label{fig:heatmap_lt_e_z_single_photons_varphi}
\end{figure}

\begin{figure}[h!]
	\centering
	\includegraphics[width=\linewidth]{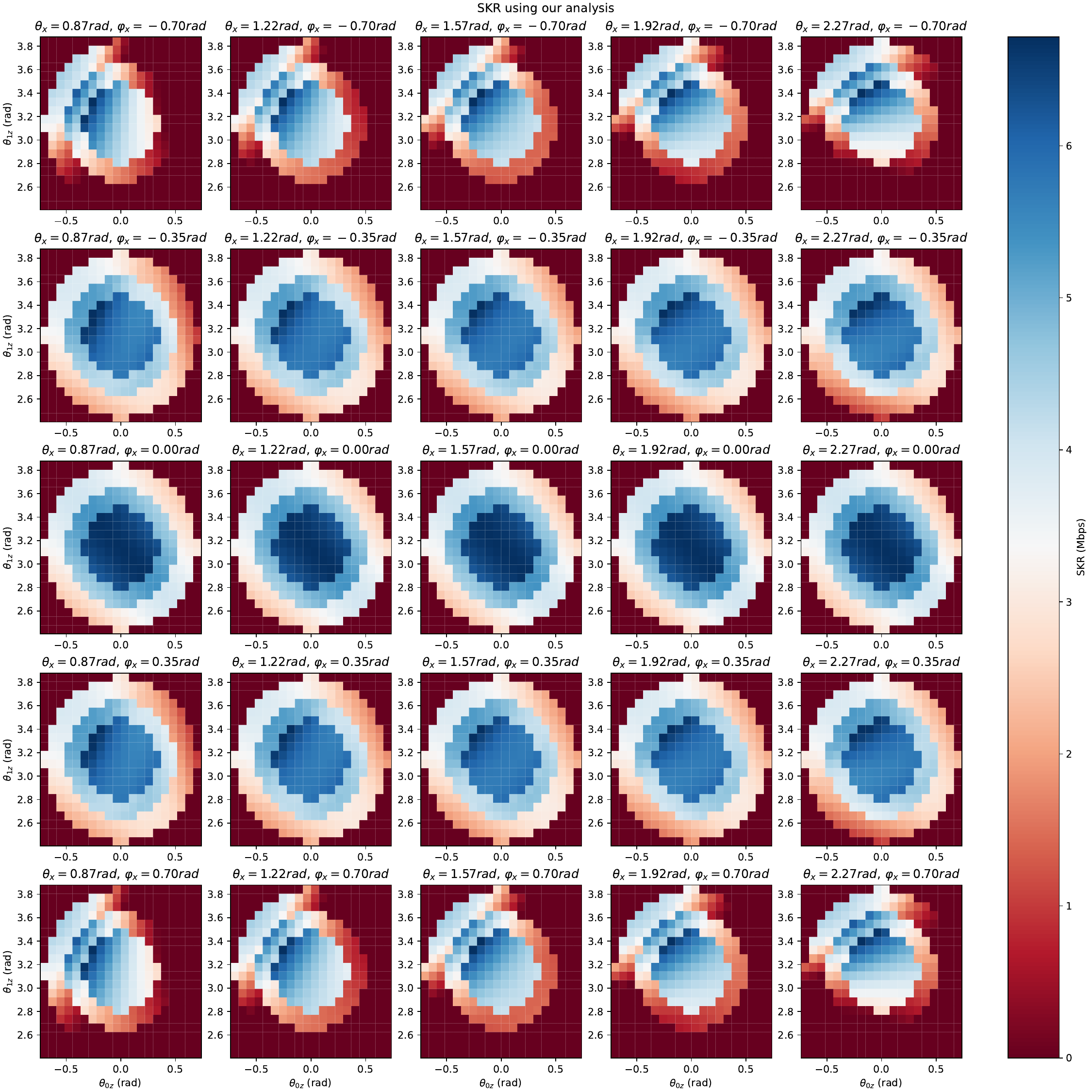}
	\caption{Simulation results using our analysis and considering single-photons. The probabilities of basis choice are fixed. The color indicates the SKR, while the X (Y) coordinates indicate $\theta_{0z(1z)}$.}
	\label{fig:heatmap_lt_skr_single_photons_varphi}
\end{figure}

In \cref{fig:heatmap_lt_e_z_single_photons_varphi}, simulated phase error rates are plotted using this analysis. 
In each plot, the  the X (Y) coordinates indicate $\theta_{0z}$  ($\theta_{1z}$) for a fixed imperfect X state. 
From left-most to the right-most plot, $\theta_{0x}$ is varied from 0.87 to 2.27 radians.
From the upper-most to lower-most plot, $\varphi_{0x}$ is also varied from -0.7 to 0.7 radians.
If we look at the plot in the center, where the 0X state is perfect, we can see that the phase error rate stays minimal along the axis in which $\delta \theta_{0Z}=\delta\theta_{1Z}$ even as the imperfections increase significantly.
The reason is that this is also the condition in which the X states of the virtual protocol are perfect.
Conversely, along the axis in which $\delta \theta_{0Z}=\delta\theta_{1Z}$, we  see that the phase error rate increases as the absolute value of the imperfections also increases, without a preference for a specific sign.
This also makes sense, since in this case, the virtual X states start to be more misaligned with respect to the ideal ones.
The corresponding simulated SKRs are plotted in \cref{fig:heatmap_lt_skr_single_photons_varphi}.

\subsubsection{Calculations of Bloch vectors and probabilities of emitting the virtual states}\label{section:bloch vectors}
In this subsection, we show the explicit calculations of the Bloch vectors and probabilities of emitting virtual states from the state characterization.
Let us start by writing the imperfect qubit states as
\begin{equation}
	\ket{\phi_{j\beta}}=\cos\frac{\theta_{j\beta}}{2}\ket{0}+\sin\frac{\theta_{j\beta}}{2}\ket{1} \label{eq:14}
\end{equation}
with $j\beta \in \{0z, 1z, 0x\}$. 
Here, $\theta_{j\beta}$ is the polar angle in the Bloch sphere of the emitted states.
We have set the azimuthal angle to $\varphi=0$ because we consider the states to lie on the X-Z plane of the Bloch sphere. 
Trivially, the Bloch vector of such states are given by
\begin{equation}
	(P_X^{j\beta}, 0, P_{Z}^{j\beta}) = (\sin\theta_{j\beta}, 0, \cos\theta_{j\beta}). \label{eq:20}
\end{equation}

Let us now calculate the probability of emitting the virtual states $\sigma_{B;jx, vir}$. 
Firstly, note that
\begin{equation}
	\ket{\Psi_Z}_{AB}=\frac{1}{2}\left[\ket{+}_A\left(\ket{\phi_{0z}}_B+\ket{\phi_{1z}}_B\right)+\ket{-}_A\left(\ket{\phi_{0z}}_B-\ket{\phi_{1z}}_B\right)\right] \label{eq:12}
\end{equation}
Then, \cref{eq:4} may be rewritten as
\begin{equation}
	\sigma_{B;jx, vir} =\frac{1}{4}\left[\ket{\phi_{0z}}\bra{\phi_{0z}}_B+ \ket{\phi_{1z}}\bra{\phi_{1z}}_B +(-1)^j\left(\ket{\phi_{1z}}\bra{\phi_{0z}}_B + \ket{\phi_{0z}}\bra{\phi_{1z}}_B\right)\right] . \label{eq:13}
\end{equation}
The probability of emitting this state is its trace, which is given by
\begin{equation}
	Tr[\sigma_{B;jx, vir}] = \frac{1}{2}\left[1 + (-1)^{j}\cos\left(\frac{\theta_{0z}-\theta_{1z}}{2}\right)\right]. \label{eq:15}
\end{equation}

Finally, we calculate the Bloch vectors $(P_X^{jx, vir}, 0, P_Z^{jx, vir})$ with the expectation value of $\sigma_{x,z}$ of the normalized virtual states. 
Defining  $N:=1/Tr[\sigma_{B;jx, vir}]$ as the normalization factor, the X and Z components are calculated as
\begin{eqnarray}
    P_X^{jx, vir} = NTr[\sigma_{B;jx, vir}\sigma_x] = \frac{N}{4}\left(
	\sin\theta_{0z} + \sin\theta_{1z} + 2(-1)^j\sin\frac{\theta_{0z}+\theta_{1z}}{2}
	\right), \label{eq:17} \\
    P_Z^{jx, vir} = NTr[\sigma_{B;jx, vir}\sigma_z]  = \frac{N}{4}\left[
	\cos\theta_{0z} + \cos\theta_{1z} 
	+ 2(-1)^j\cos\left(\frac{\theta_{0z}+\theta_{1z}}{2}\right)
	\right]. \label{eq:19}
\end{eqnarray}

\clearpage
\subsection{Reduction to perfect state preparation case}\label{section:reduction perfect}

As a sanity check, let us calculate the phase error rate with perfect state preparation. This should be the same as in \cite{rusca_security_2018}, since we are following the same method of estimating the experimentally unavailable statistics.

Using \cref{eq:solution_qs}, we get
\begin{eqnarray}
    &q_{1x|Id} = \frac{2}{p_Z^Ap_X^B}\left(Y_{t1, 0z}^{(exp)} + Y_{t1, 1z}^{(exp)}\right) ,\label{eq:43}\\
    &q_{1x|z} = \frac{2}{p_Z^Ap_X^B}\left(Y_{t1, 0z}^{(exp)} - Y_{t1, 1z}^{(exp)}\right), \label{eq:44} \\
    &q_{1x|x} = \frac{2}{(1-p_Z^A)p_X^B}Y_{t1,0x}^{(exp)} - \frac{2}{p_Z^Ap_X^B}\left(Y_{t1, 0z}^{(exp)} + Y_{t1, 1z}^{(exp)}\right), \label{eq:45}\\
    &q_{0x|Id} = \frac{4}{p_Z^Ap_X^B}\left[Y_{side,0z}^{(exp)} + Y_{side,1z}^{(exp)} -\frac{1}{2}(Y_{t1,0z}^{(exp)}+Y_{t1,1z}^{(exp)})\right], \label{eq:49} \\
    &q_{0x|z} = \frac{4}{p_Z^Ap_X^B}\left[Y_{side,0z}^{(exp)} - Y_{side,1z}^{(exp)} +\frac{1}{2}(Y_{t1,1z}^{(exp)}-Y_{t1,0z}^{(exp)})\right], \label{eq:50} \\
    &q_{0x|x} = \frac{4}{(1-p_Z^A)p_X^B}\left[Y_{side,0x}^{(exp)} -\frac{1}{2}Y_{t1,0x}^{(exp)}
	\right] - \frac{4}{p_Z^Ap_X^B}\left[Y_{side,0z}^{(exp)} + Y_{side,1z}^{(exp)} -\frac{1}{2}(Y_{t1,0z}^{(exp)}+Y_{t1,1z}^{(exp)})\right]. \label{eq:51}
\end{eqnarray}
Here, we have used the definition $Y_{side,j\beta}^{(exp)}:=Y_{t0,j\beta}^{(exp)}+Y_{t2,j\beta}^{(exp)}$.

The yields of the virtual protocol are calculated with \cref{eq:36,eq:37,eq:38,eq:39} as
\begin{eqnarray}
    &Y_{0x,0x}^{(Z)} = \frac{1}{2}p_X^{B} \left(
	q_{0x|Id} + q_{0x|x} \right), \label{eq:52} \\
    &Y_{1x,0x}^{(Z)} = \frac{1}{2}p_X^{B} \left(
	q_{1x|Id} +q_{1x|x} \right), \label{eq:53} \\
    &Y_{0x,1x}^{(Z)} = \frac{1}{2}p_X^{B} \left(
	q_{0x|Id} - q_{0x|x} \right), \label{eq:54} \\
    &Y_{1x,1x}^{(Z)} = \frac{1}{2}p_X^{B} \left(
	q_{1x|Id} - q_{1x|x} \right).  \label{eq:55}
\end{eqnarray}
Using the found $q_{sx|t}$ yields
\begin{eqnarray}
    &Y_{0x,0x}^{(Z)} = \frac{2}{1-p_Z^A}\left[Y_{side,0x}^{(exp)} -\frac{1}{2}Y_{t1,0x}^{(exp)}
	\right],  \label{eq:56} \\
    &Y_{1x,0x}^{(Z)} = \frac{1}{1-p_Z^A}Y_{t1,0x}^{(exp)},  \label{eq:57} \\
    &Y_{0x,1x}^{(Z)} = \frac{4}{p_Z^A}\left[Y_{side,0z}^{(exp)} + Y_{side,1z}^{(exp)} -\frac{1}{2}(Y_{t1,0z}^{(exp)}+Y_{t1,1z}^{(exp)})\right] - \frac{2}{1-p_Z^A}\left[Y_{side,0x}^{(exp)} -\frac{1}{2}Y_{t1,0x}^{(exp)}
	\right],  \label{eq:58} \\
    &Y_{1x,1x}^{(Z)} =  \frac{2}{p_Z^A}\left(Y_{t1, 0z}^{(exp)} + Y_{t1, 1z}^{(exp)}\right) - \frac{1}{1-p_Z^A}Y_{t1,0x}^{(exp)} . \label{eq:59}
\end{eqnarray}

Before plugging these values in \cref{eq:single-photon_phase_error_rate}, let us first define the conditional probability $Y_{ts|j\beta}=Y_{ts,j\beta}^{(\beta)}/p_{j\beta}$ , with $p_{j\beta}$ being the probability of preparing the state $j_{\beta} \in \{0Z, 1Z, 0X\}$. Then, the sum of \cref{eq:56,eq:57,eq:58,eq:59} yields
\begin{equation}
	\sum_{s,j}Y_{sx, jx}^{(Z)} = 2(Y_{side|0z}+Y_{side|1z}) = 2Y_{side|Z}. \label{eq:60}
\end{equation}
Then,
\begin{equation}
	Q_X = \frac{Y_{t1|0x}}{2Y_{side|Z} } +  \frac{Y_{side|Z}-Y_{side|0x}-\frac{1}{2}(Y_{t1|0z}+Y_{t1|1z}-Y_{t1|0x})}{Y_{side|Z}}. \label{eq:61}
\end{equation}
Here, $Q_X$ can be negative. We ensure that $Y_{0x,1x}^{(Z)}$ is non-negative by writing  
\begin{equation}
	Q_X = \frac{Y_{t1|0x}}{2Y_{side|Z} } + \mathcal{M}\left\{1+ \frac{\frac{1}{2}(Y_{t1|0x}-Y_{t1|0z}-Y_{t1|1z})-Y_{side|0x}}{Y_{side|Z}} \right\}, \label{eq:62}
\end{equation}
where $\mathcal{M}\{x\} =max(0,x)$. As expected, this is the same results as in \cite{rusca_security_2018}.

\begin{figure}[h!]
	\centering
	\includegraphics[width=\linewidth]{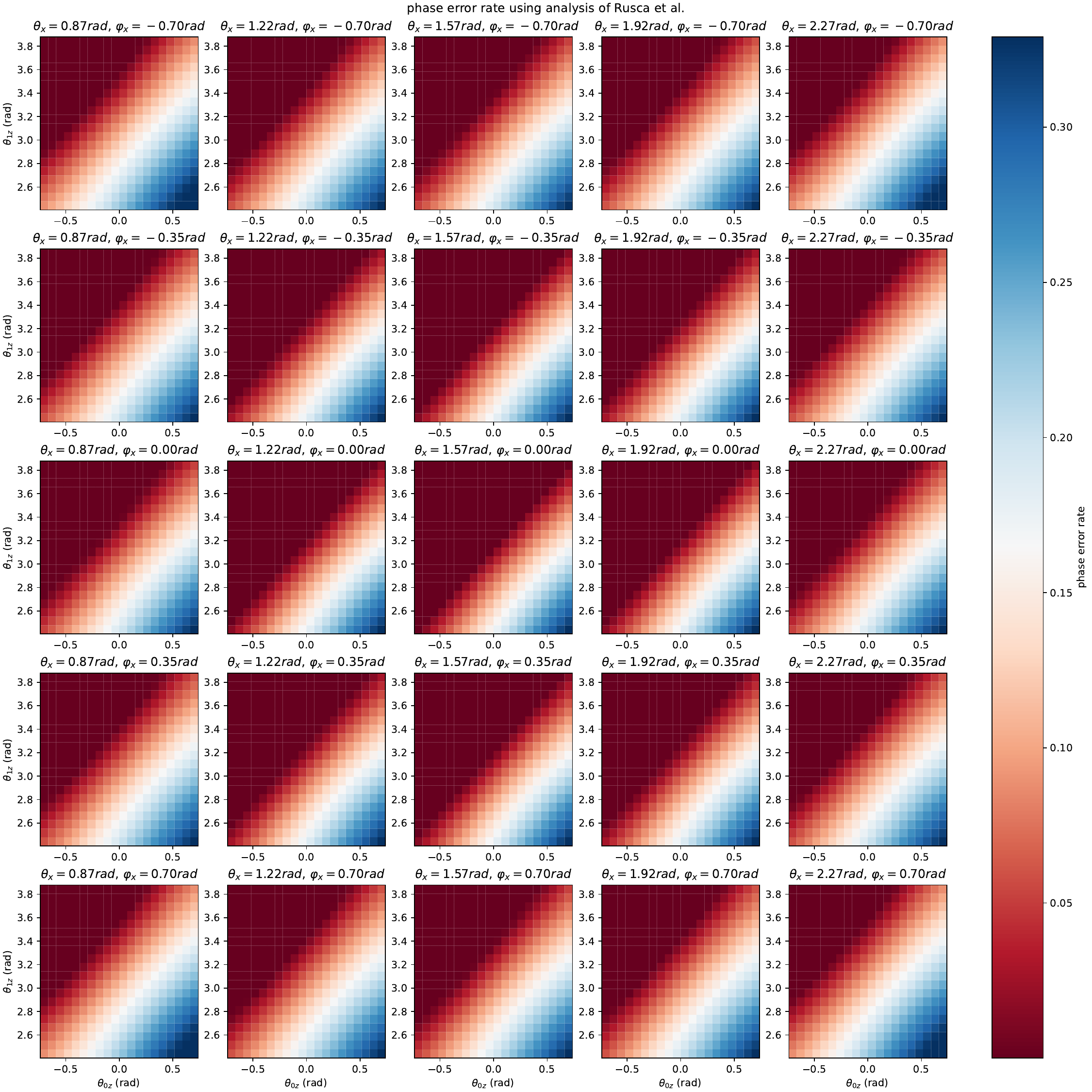}
	\caption{Simulation results using the security proof of \cite{rusca_security_2018} considering single-photons. The probabilities of basis choice are fixed. The color indicates the phase error rate, while the X (Y) coordinates indicate $\theta_{0z(1z)}$.}
	\label{fig:heatmap_rusca_e_z_single_photons_varphi}
\end{figure}

\begin{figure}[h!]
	\centering
	\includegraphics[width=\linewidth]{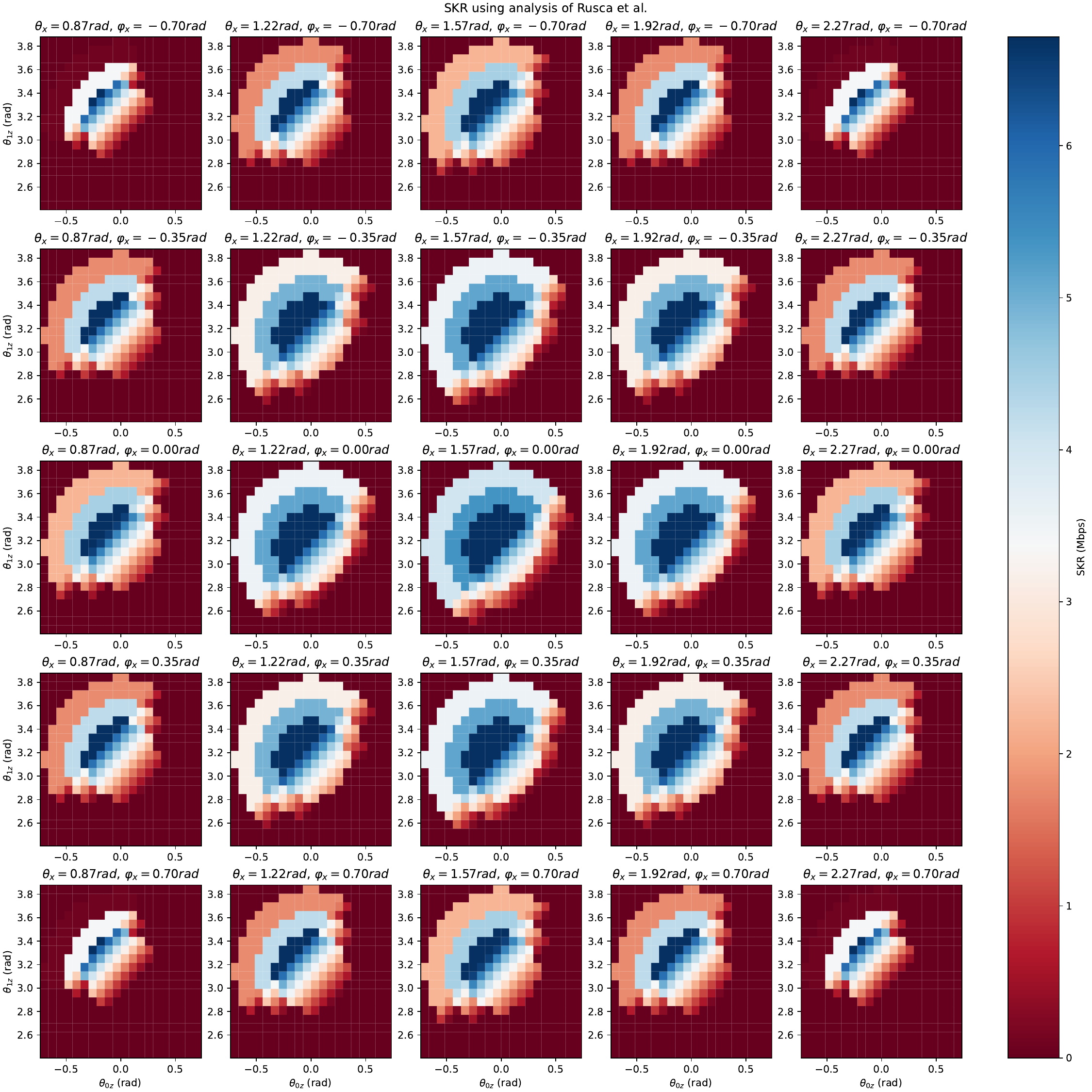}
	\caption{Simulation results using the security proof of \cite{rusca_security_2018} considering single-photons. The probabilities of basis choice are fixed. The color indicates the SKR, while the X (Y) coordinates indicate $\theta_{0z(1z)}$.}
	\label{fig:heatmap_rusca_skr_single_photons_varphi}
\end{figure}

We use \cref{eq:62} to simulate again the phase error rate in the same cases as in \cref{fig:heatmap_rusca_e_z_single_photons_varphi} but assuming perfect state preparation.
Remarkably, the imperfections on the 0X state seem to have no impact on the results.
Moreover, we observe that the phase error rate is minimized when $\delta\theta_{1Z}>\delta\theta_{0Z}$.
The SKR is plotted in \cref{fig:heatmap_rusca_skr_single_photons_varphi}.

\clearpage

\subsection{One decoy analysis}
The phase error rate so far was calculated assuming single photons. 
Following the finite-key analysis described in \cite{rusca_finite-key_2018}, this fictitious error rate should be upper-bounded for single photons. We denote this bound with $\overline{Q}_X$. 
Then, the phase error rate, $\phi_Z$, is given by \cref{eq:phase_error_rate_decoy}.
To calculate it, one needs to estimate bounds for the vacuum and single-photon events.

\subsubsection{Estimation of vacuum events}

As in \cite{rusca_security_2018}, we estimate the vacuum events using the \quotes{empty} time bins. We start by defining $n_{empty}$ as the errors of the Z states, that is,
\begin{equation} 
	n_{empty} = n(X2, 0Z) + n(X0, 1Z). \label{eq:67}
\end{equation}
Then, we estimate the vacuum events corresponding to Alice sending the state $a \in \{0Z, 1Z, 0X\}$ and Bob having a detection in $ti$ as
\begin{equation}
	v^{(exp)}_{ti,a} = \frac{p_a^{A}}{p_Z^{A}}n_{empty} \quad \text{with} \quad a\in\{0z, 1z, 0x\}. \label{eq:73}
\end{equation}

\subsubsection{Bounds for the virtual protocol}

One can write the events of the virtual protocol as
\begin{equation}
	n^{(vir)}_{sx, jx} = \sum_{i,a} M^{i,a}_{s, j}n^{(exp)}_{ti, a} = \sum_{i,a} M^{i,a}_{s, j}\left(n^{(exp)}_{ti, a, \mu0}+n^{(exp)}_{ti, a, \mu1} \right). \label{eq:69}
\end{equation}
Here,  $M_s=AB^{-1}C_s$, where the matrices $C_s$ are defined from \cref{eq:definition_b1_and_qs,eq:definition_b0} as \newline $\mathbf{b_s}=\sum_{i=0,1,2}\sum_{j\beta=0Z, 1Z, 0X}C_s^{i,j\beta}\mathbf{Y_{ti, j\beta}^{(exp)}}$. 
Then, one can define the counts of that event in which the intensity of the state is $\mu_k$ as
\begin{equation}
	n^{(vir)}_{sx, jx, \mu k} = \sum_{i,a} M^{i,a}_{s, j}n^{(exp)}_{ti, a, \mu k} \label{eq:70}
\end{equation}
and bound directly the virtual single-photon events related to this quantity. 
The lower bound is given by
\begin{equation}
	\underline s^{(vir)}_{sx, jx}=\frac{\tau_1\mu_0}{\mu_1(\mu_0-\mu_1)}\left(
	n^{(vir)-}_{sx, jx, \mu1} - \frac{\mu_1^2}{\mu_0^2}n^{(vir)+}_{sx, jx, \mu0}
	- \frac{(\mu_0^2-\mu_1^2)}{\mu_0^2}\frac{\overline{v^{(vir)}}_{sx, jx}}{\tau_0}
	\right), \label{eq:71}
\end{equation}
where 
\begin{equation}
	n_{m,n,\mu k}^{\pm} = \frac{e^{\mu_k}}{p_{\mu k}}\left( n_{m,n,\mu k} \pm \delta(n_{m,n}, \frac{\epsilon_{sec}}{19}) \right),  \label{eq:75}
\end{equation}
and $\delta(n, \epsilon) = \sqrt{n\log(\epsilon^{-1})/2}$ is the finite-key correction given by Hoeffding's inequality, $\tau_1$ is the total probability to send a single photon and $\overline{v^{(vir)}}_{sx, jx}$ is an upper bound on the vacuum events given by
\begin{equation}
	\overline{v^{(vir)}}_{sx, jx} =  \sum_{i,a}M^{i,a}_{s, j}v_{ti,a}^{(exp)} + \delta\left(n^{(vir)}_{sx,jx}, \frac{\epsilon_{sec}}{19}\right), \label{eq:72}
\end{equation}

Finally, the upper bound on single-photon events is given by
\begin{equation}
	\overline s^{(vir)}_{sx, jx}=\tau_{1}\frac{n^{(vir)+}_{sx, jx, \mu0} - n^{(vir)-}_{sx, jx, \mu1}}{\mu_0 - \mu_1}. \label{eq:74}
\end{equation}

\clearpage

\clearpage

\bibliographystyle{ieeetr}

\bibliography{bibliography.bib}

\end{document}